\documentclass[aps,pra,superscriptaddress,amsmath,amssymb,preprintnumbers,showpacs,twocolumn,reprint]{revtex4-1}
\usepackage{graphicx}
\usepackage{amssymb}
\usepackage{qcircuit}
\usepackage{braket}
\usepackage{url}
\usepackage{amsmath}
\usepackage{dcolumn}
\usepackage{bm}
\usepackage{braket}
\usepackage{here}
\usepackage[caption=false]{subfig}
\usepackage{color}

\newcommand{\mr}[1]{\mathrm{#1}}

\begin{document}

\title{Variational quantum simulations of stochastic differential equations}

\author{Kenji Kubo}
\email{kenjikun@mercari.com}
\affiliation{R4D, Mercari Inc., Roppongi Hills Mori Tower 18F, 6-10-1, Roppongi, Minato-ku, Tokyo 106-6118, Japan}
\affiliation{Graduate School of Engineering Science, Osaka University, 1-3, Machikaneyama, Toyonaka, Osaka 560-8531, Japan }
\author{Yuya O. Nakagawa}
\email{nakagawa@qunasys.com}
\affiliation{QunaSys Inc., Aqua Hakusan Building 9F, 1-13-7 Hakusan, Bunkyo, Tokyo 113-0001, Japan}
\author{Suguru Endo}
\email{suguru.endou.uc@hco.ntt.co.jp}
\affiliation{NTT Secure Platform Laboratories, NTT Corporation, Musashino, Tokyo 180-8585, Japan}
\author{Shota Nagayama}
\email{shota.nagayama@mercari.com}
\affiliation{R4D, Mercari Inc., Roppongi Hills Mori Tower 18F, 6-10-1, Roppongi, Minato-ku, Tokyo 106-6118, Japan}
\date{\today}

\begin{abstract}
Stochastic differential equations (SDEs), which models uncertain phenomena as the time evolution of random variables, are exploited in various fields of natural and social sciences such as finance.
Since SDEs rarely admit analytical solutions and must usually be solved numerically with huge classical-computational resources in practical applications, there is strong motivation to use quantum computation to accelerate the calculation.
Here, we propose a quantum-classical hybrid algorithm that solves SDEs based on variational quantum simulation (VQS).
We first approximate the target SDE by a trinomial tree structure with discretization and then formulate it as the time-evolution of a quantum state embedding the probability distributions of the SDE variables.
We embed the probability distribution directly in the amplitudes of the quantum state while the previous studies did the square-root of the probability distribution in the amplitudes.
Our embedding enables us to construct simple quantum circuits that simulate the time-evolution of the state for general SDEs.
We also develop a scheme to compute the expectation values of the SDE variables and discuss whether our scheme can achieve quantum speed-up for the expectation-value evaluations of the SDE variables.
Finally, we numerically validate our algorithm by simulating several types of stochastic processes.
Our proposal provides a new direction for simulating SDEs on quantum computers.%
\end{abstract}

\maketitle

\section{Introduction \label{sec: intro}}
Stochastic differential equations (SDEs), which describe the time evolution of random variables, are among the most important mathematical tools for modeling uncertain systems in diverse fields, such as finance \cite{Shreve2004}, physics \cite{Kampen2011}, and biology \cite{Wilkinson2009}.
From the expectation values of the simulated random variables, we can often extract information about the system of interest.
Since the expectation values rarely admit analytical solutions, they are usually obtained by numerical methods such as the Monte Carlo method~\cite{Metropolis1949}.
However, those numerical methods incur high computational costs, especially in high-dimensional problems such as the SDEs of financial applications~\cite{Black1973, Longstaff2001, Kloeden1992}. Therefore, a method that can speed-up SDE simulations is urgently demanded.

Such a speed up can be achieved on quantum computers.
Throughout the past decade, technological developments have realized a primitive form of quantum computers called noisy intermediate-scale quantum (NISQ) devices~\cite{Preskill2018}, which can handle problems that are intractably large for classical computers~\cite{Arute2019}.
NISQ devices can operate only a few tens to hundreds of qubits without error correction, so they cannot run quantum algorithms requiring deep and complicated quantum circuits.
Although quantum algorithms are expected to outperform classical ones on specific computing tasks~\cite{Grover96, Shor94, Harrow09, Nielsen2010}, they usually exceed the capability of NISQ devices. Accordingly, NISQ devices have been leveraged with heuristic algorithms that solve real-world problems.
For example, in quantum chemistry and condensed matter physics, the variational quantum eigensolver (VQE) algorithm~\cite{Peruzzo2014, Kandala2017} can calculate the ground-state energies of given Hamiltonians~\cite{McArdle2020,cao2019quantum}.
Another example is quantum machine learning with variational quantum circuits~\cite{Mitarai2018, Schuld2019, Farhi2018, Cong2019}.
Both algorithms variationally optimize the tuneable classical parameters in quantum circuits, so the speedups of the computation over classical computers and the precision of the obtained results are not guaranteed in general.

Several quantum-computing-based methods obtain the expectation value of a function that takes an SDE solution as its argument. However, all of these methods require prerequisite knowledge of the SDE solution.
In \cite{Fontanela2019}, the partial differential equation describing the time evolution of the expectation value was simulated by a variational quantum computation, which requires pre-derivation of the partial differential equation of the expectation value.
In \cite{Rebentrost2018} and \cite{Zoufal2019}, the probability distribution of the SDE solution was embedded in the quantum state, and the expectation value was calculated by a quantum amplitude estimation algorithm (QAE). In this case, the probability distribution of the SDE solutions must be known in advance.
As the solution to the SDE is not found, the partial differential equation of the expectation value must also be derived, or the SDE solved beforehand.

In this study, to solve an SDE with quantum algorithms, we apply a tree model approximation~\cite{P.Boyle1986}, and hence obtain a linear differential equation describing the probability distribution of SDE solutions. This differential equation is then solved by a variational quantum simulation (VQS)~\cite{Li2017, McArdle2019, Yuan2019, Endo2020,endo2020hybrid}. Note that linear differential equations can be solved by a quantum linear solver algorithm (QLSA)~\cite{Harrow09,Berry2014,Berry2017}, which is expected to be quantum-accelerated.
However, the QLSA requires a large number of ancilla qubits and deep circuits and is possibly executable only on quantum computers with error correction.
Our proposed method possesses several desirable features. First, the probability distribution is simulated by the tree-model approximation, so the model requires only the SDE. No prior knowledge of the probability distribution or expectation value is required. Therefore, our method is applicable to more general SDEs than previous methods. Second, once the VQS is performed, the variational parameters are obtained as classical information, and the probability distribution of the simulation results can be used to compute various expectation values. We can also compute path-dependent expectation values because the time series of the probability distribution is obtained. Third, the algorithm is less resource-intensive than the QLSA.
Since VQS is a variational algorithm, it is difficult to estimate the exact computational cost, but VQS requires only a few ancilla qubits and calculates the expectation value for relatively shallow unitary gates at each time step.
The number of qubits and the depth of the circuit are expected to be much smaller than QLSA.
As our method uses a new scheme for embedding probability distributions in quantum states, the method for computing expectation values is also new. We additionally found that the expectation values are more simply determined by our method than by the QAE. The proposed method facilitates the application of SDEs in quantum computing simulations and is expected to impact various scientific fields.

The remainder of this paper is organized as follows.
Section~\ref{sec: reviews} reviews the trinomial tree-model approximation and the VQS, before introducing our method.
Our main theoretical results are contained in Secs.~\ref{sec: sde_by_vqs}, \ref{sec: expectation}.
Section~\ref{sec: sde_by_vqs} proposes a VQS-based method that simulates the dynamics of the probability distribution of the stochastic process in the trinomial tree model.
The quantum circuits and operators that perform the VQS are also constructed in this section.
Section~\ref{sec: expectation} calculates the expectation value of the random variable using the state obtained by simulating SDE with the VQS.
Section~\ref{sec: discussion} discusses the advantages of our method and compares them with previous studies.
Section~\ref{sec: numerics} numerically evaluates our algorithm on two SDE prototypes: the geometric Brownian motion and the Ornstein-Uhlenbeck process.
Conclusions are presented in Section \ref{sec: conclusion}.
Appendix~\ref{app: complexity of expectation} analyses the complexity of calculating the expectation value, and Appendix~\ref{app: multi-variables} generalizes our result to a multiple-variable process.
Appendix~\ref{app: error from piecewise polynomial approximation} evaluates the error of expectation values from piecewise polynomial approximation.%
\section{Preliminaries \label{sec: reviews}}
This section reviews the main ingredients of this paper: the trinomial tree-model approximation of the SDE~\cite{P.Boyle1986} and the VQS algorithm~\cite{Li2017, McArdle2019, Yuan2019, Endo2020}.
In Sec.~\ref{sec: sde_by_vqs}, we combine both ingredients into a method that simulates the SDE by the VQS.

\subsection{Trinomial tree-model approximation of the stochastic differential equation}
\begin{figure}
  \centering
  \includegraphics[width=25em]{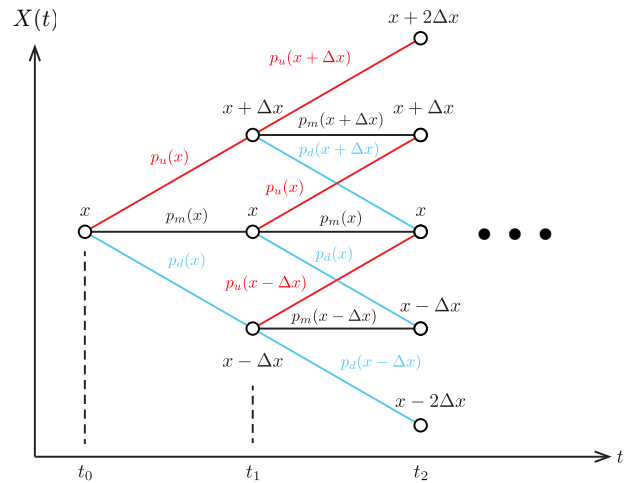}

  \caption{Lattice of the trinomial tree model. Nodes (circles) at $(t, x)$ represent the events in which $X(t)$ takes the value $x$.
    Edges represent the transition probabilities between the nodes.
    The stochastic process starts at node $(t_0, x_0)$ and ``hops'' to the other nodes depending on the transition probabilities.}
  \label{fig:tree_model}
\end{figure}

Let us consider a random variable $X(t)$ taking values on an interval $I \subset \mathbb{R}$. We refer to $I$ as an event space.
The SDE of a single process $\{X(t)\}_{t \in [0, T]}$, which is a time-series of random variables from $t=0$ to $t=T$, is defined as~\cite{Shreve2004}
\begin{equation}
  dX(t) = \mu(X(t), t)dt + \sigma(X(t), t)dW, \: X(0) = x_{\mr{ini}} \in I, \label{eq: general SDE def}
\end{equation}
where $\mu(X(t), t), \sigma(X(t), t)$ are real valued functions of time $t$ and the variable $X(t)$, and $W$ denotes the Brownian motion.
In the main text, our proposal is applied to a single process (extensions to multi-variables cases are described in Appendix~\ref{app: multi-variables}).

The tree model numerically simulates the time evolution of an SDE.
Let us consider an SDE simulation of the process with event space $[0,x_{\max}]$ from $t=0$ to $t=T$.
We discretize the time as $t_i \equiv i\Delta t \: (i=0,1,\dots, N_t)$ and the event space as $x_i \equiv i\Delta x \: (i=0,1,\dots, N_x)$, where $N_t\Delta t = T$ and $N_x\Delta x = x_{\max}$.
In this discretization scheme, we define a $(N_x+1) \times (N_t+1)$ lattice on which each node $(i,j)$ is associated with a probability $\mr{Prob}[X(t_j) = x_i]$ and each edge represents a transition between two nodes, as shown in Fig.~\ref{fig:tree_model}.
Here, we adopt the trinomial tree model, which has three transition probabilities as follows:
\begin{eqnarray*}
  p_u(x, t) &=& \mr{Prob}[X(t+\Delta t)=x+\Delta x \,|\, X(t)=x], \\
  p_d(x, t) &=& \mr{Prob}[X(t+\Delta t)=x-\Delta x \,|\, X(t)=x], \\
  p_m(x, t) &=& \mr{Prob}[X(t+\Delta t)=x \,|\, X(t)=x].
\end{eqnarray*}
These probabilities were chosen to reproduce the first and second moment (mean and variance, respectively) of the random variable $X(t)$ in Eq.~\eqref{eq: general SDE def}.
Following the Euler-Maruyama method~\cite{Kloeden1992}, the SDE is discretized as
\begin{equation}
  X(t_{j+1}) - X(t_j) = \mu(X(t_j), t)\Delta t + \sigma(X(t_j),t)\sqrt{\Delta t} z,
\end{equation}
where $z \sim N(0,1)$ and $O(\Delta t^2)$ terms are ignored.
The conditional expectation value and variance are respectively expressed as
\begin{eqnarray*}
  E[X(t_{j+1})-X(t_j)|X(t_j)=x] &=&\mu(x,t_j)\Delta t, \\
  \mr{Var}[X(t_{j+1})-X(t_j)|X(t_j)=x] &=& \sigma^2(x,t_j)\Delta t.
\end{eqnarray*}
The corresponding moments on the trinomial tree model are
\begin{eqnarray*}
  &&E[X(t_{j+1})-X(t_j)|X(t_j)=x_i]  \\
  &=& (p_u(x_i,t_j) - p_d(x_i, t_j)) \Delta x, \\
  &&\mr{Var}[X(t_{j+1})-X(t_j)|X(t_j)=x_i]  \\
  &=& (p_u(x_i,t_j) + p_d(x_j, t_j)) \Delta x^2.
\end{eqnarray*}
Equating these moments and considering the normalization condition $p_u(x, t) + p_m(x, t) + p_d(x, t) = 1$, we obtain
\begin{eqnarray}
  p_u(x_i, t_j) &=& \frac{1}{2}\left(\frac{\sigma^2(x_i,t_j)}{\Delta x^2}+\frac{\mu(x_i,t_j)}{\Delta x}\right)\Delta t, \label{eq: p_u} \\
  p_d(x_i, t_j) &=& \frac{1}{2}\left(\frac{\sigma^2(x_i,t_j)}{\Delta x^2}-\frac{\mu(x_i,t_j)}{\Delta x}\right)\Delta t, \label{eq: p_d} \\
  p_m(x_i, t_j) &=& 1-\frac{\sigma^2(x_i,t_j)}{\Delta x^2} \Delta t. \label{eq: p_m}
\end{eqnarray}
In summary, the trinomial tree-model approximates the original SDE by discretizing it on the lattice and setting the transition probabilities between the nodes to reproduce the first and the second moments of the process.

The trinomial tree model simulates the SDE as follows.
First, the closest value to $x_\mr{ini}$ in $\{x_i\}_{i\in[0,N_x]}$ is set to $x_{i_0}$, and the probabilities are set as $\mr{Prob}[X(t_0) = x_{i_0}] = 1, \mr{Prob}[X(t_0) = x_{i\neq i_0}] = 0$.
Next, the probability distribution of $X(t_1 = \Delta t)$ is calculated using the transition probabilities given by Eqs.~\eqref{eq: p_u}\eqref{eq: p_d}\eqref{eq: p_m}.
Repeating this step for $X(t_j) (j=2,3,...,N_t-1)$ yields all probabilities $\mr{Prob}[X(t_j)=x_i]$ at node $(i,j)$, from which any properties related to the process $X(t)$, such as the expectation values of $X(T)$ under some function $f$, $E[f(X(T))]$, can be determined.
In option-pricing financial problems, the nodes of the tree model denote the prices of the option, and the problems are sometimes to be solved in the backward direction from time $t$. In such cases, the boundary condition is set at $t=T$.

\subsection{Variational quantum simulation (VQS)}
This subsection introduces the VQS algorithm~\cite{Li2017, McArdle2019, Endo2020, Yuan2019}, a quantum-classical hybrid algorithm that simulates both unitary and non-unitary time evolution with possibly shallow quantum circuits. Therefore, the VQS algorithms is especially suitable for NISQ devices.

We are interested in the non-unitary time evolution of an unnormalized quantum state $\ket{\tilde{\psi}(t)}$ on an $n$-qubit system, defined as
\begin{equation} \label{eq: time evolution}
  \frac{d}{dt} \ket{\tilde{\psi}(t)} = L(t) \ket{\tilde{\psi}(t)},
\end{equation}
where $L(t)$ is a time-dependent (possibly non-Hermitian) linear operator.
To simulate the dynamics of $\ket{\tilde{\psi}(t)}$, let us introduce the following ansatz quantum state $\ket{\tilde{v}(\bm{\theta}(t))}$:
\begin{eqnarray} \label{eq: unnoraml state def}
  \ket{\tilde{v}(\bm{\theta}(t))} \equiv\alpha(t) R(\bm{\theta}_1(t)) \ket{0}
\end{eqnarray}
where $\alpha(t)$ is a real number, $\bm{\theta}(t) \equiv (\alpha(t), \bm{\theta}_1(t)) \equiv (\alpha(t), \theta_1(t), \ldots, \theta_M(t))$ are variational parameters of the ansatz, $\ket{0}$ is some reference state, and  $R(\bm{\theta}_1) \equiv R_1(\theta_1)R_2(\theta_2)\cdots R_M(\theta_M)$ is a product of $M$ parametric circuits (unitaries) composed of one parametric rotation gates $e^{i\theta_k G_{k}} (G_{k}^\dag = G_{k})$. The gates depend on their parameters and on other non-parametric gates.
In particular, $G_{k}$ is assumed as a multi-qubit Pauli gate $\{I, X, Y, Z\}^{\otimes n}$.

The VQS algorithm maps the dynamics of the quantum state, Eq.~\eqref{eq: time evolution}, to those of the variational parameters $\bm{\theta}(t)$ of the ansatz.
The mapping is performed by McLachlan's variational principle~\cite{Mclachlan1964}
\begin{eqnarray}
  \min_{\bm{\theta}(t)} \left\| \frac{d}{dt}\ket{\tilde{v}(\bm{\theta}(t)} - L(t) \ket{\tilde{v}(\bm{\theta}(t)} \right\|,
\end{eqnarray}
where $\|\ket{\varphi}\| \equiv \sqrt{\braket{\varphi|\varphi}}$.
This equation reduces to an Euler-Lagrange equation,
\begin{equation}\label{eq: Euler-lagrange}
  \sum_{j=0}^M M_{k,j}\dot{\theta}_j(t) = V_k,
\end{equation}
for $k=0,\cdots,M$ where
\begin{eqnarray}
  M_{k,j} &\equiv& \Re\left(\frac{\partial \bra{\tilde{v}(\bm{\theta}(t))}}{\partial \theta_k}\frac{\partial\ket{\tilde{v}(\bm{\theta}(t))}}{\partial \theta_j}\right),  \label{eq: M matrix}\\
  V_k &\equiv& \Re\left(\frac{\partial\bra{\tilde{v}(\bm{\theta}(t))}}{\partial \theta_k} L(t) \ket{\tilde{v}(\bm{\theta}(t))} \right). \label{eq: V vector}
\end{eqnarray}
We define $\theta_0(t) \equiv \alpha(t)$ for notational simplicity.
When simulating the dynamic Eq.~\eqref{eq: time evolution}, one starts from the initial parameters $\bm{\theta}_\mr{ini}$ corresponding to the initial state $\ket{\tilde{\psi}(t=0)} = \ket{\tilde{v}(\bm{\theta}_\mr{ini})}$.
The time derivative $\dot{\bm{\theta}}(t=0)$ is calculated by Eq.~\eqref{eq: Euler-lagrange} with $\ket{\tilde{v}(\bm{\theta}_\mr{ini})}$ in Eqs.~\eqref{eq: M matrix} and~\eqref{eq: V vector}. After a small time step $\delta t$, the parameters are obtained as $\bm{\theta}(\delta t) = \bm{\theta}_\mr{ini} + \delta t \cdot \dot{\bm{\theta}}(t=0)$.
Repeating this procedure obtains the dynamics of $\bm{\theta}(t)$ and the state $\ket{\tilde{v}(\bm{\theta}(t))}$.

\begin{figure*}[t]

  \subfloat[]{
  \centering
  \mbox{
  \Qcircuit @C=2.0em @R=0.7em {
  \lstick{(\ket{0}+e^{i\theta}\ket{1})/\sqrt{2}} & \qw & \qw & \gate{X} &  \ctrl{2}  & \gate{X} &  \qw & \qw & \ctrl{2} & \gate{H} &  \meter\\
  &&\cdots&&&&\cdots&&&&\\
  \lstick{\ket{0}} &  \gate{R_N} & \qw & \gate{R_{k}} & \gate{G_{k}} &  \gate{R_{k-1}} & \qw& \gate{R_{j}} & \gate{G_{j}} &  \qw & \qw \\
  }
  }
  }
  \\
  \subfloat[]{
  \centering
  \mbox{
  \Qcircuit @C=2.0em @R=0.7em {
  \lstick{(\ket{0}+e^{i\theta}\ket{1})/\sqrt{2}} & \qw & \qw & \gate{X} &  \ctrl{2}  & \gate{X} &  \qw & \qw & \ctrl{2} & \gate{H} &  \meter\\
  &&\cdots&&&&\cdots&&&&\\
  \lstick{\ket{0}} &  \gate{R_N} & \qw & \gate{R_{k}} & \gate{G_{k}} &  \gate{R_{k-1}} & \qw& \gate{R_{1}} & \gate{U_{j}} &  \qw & \qw \\
  }
  }
  }
  \caption{
    Quantum circuits for evaluating (a) $M_{k,j}$ and (b) $V_k$.
  }
  \label{fig:circuitMV}
\end{figure*}
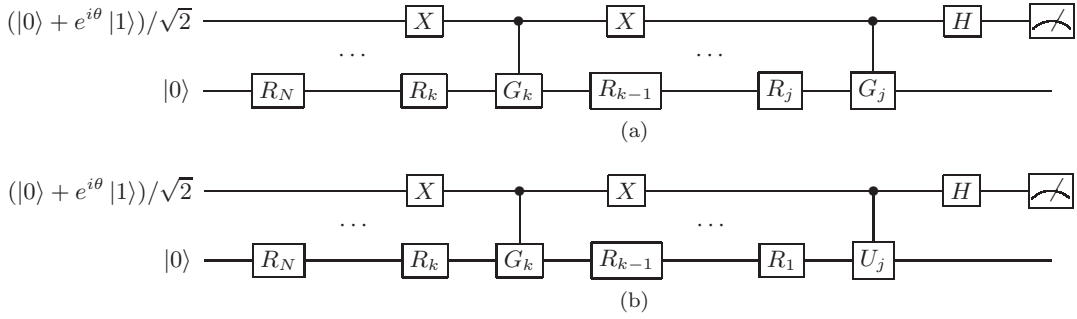

The terms $M_{k,j}$ and $V_k$ can be evaluated by the quantum circuits depicted in Fig.~\ref{fig:circuitMV}~\cite{Endo2020}.
The normalized state $\ket{v(\bm{\theta}(t))} \equiv R(\bm{\theta}_1(t)) \ket{0}$ is actually prepared on quantum computers and is multiplied by the normalization constant $\alpha(t)$ in post-processes of the result of the circuit measurements.
Decomposing the operator $L(t)$ as $L(t) = \sum_{k=1}^{k_{\rm{term}}(t)} \lambda_k U_k$, where $U_k$ is an easily-implementable unitary operator (e.g., a multi-qubit Pauli operator) and $\lambda_k$ is a complex coefficient, we must evaluate $O(M^2) + O(M k_{\rm{term}}(t))$ distinct quantum circuits.
The circuits need an ancilla qubit other than the qubits of the system of interest, along with control operations of $G_k$ and $U_k$.
Therefore, to ensure a feasible VQS algorithm, both $M, k_{\rm{term}}(t)$, and the depth of the unitaries $U_k$ must be $O(\mr{poly}(n))$.
\section{Solving stochastic differential equations by variational quantum simulation \label{sec: sde_by_vqs}}

This section presents one of our main results.
The SDE simulated by the above-described trinomial tree model is reformulated as the non-unitary dynamics of a quantum state $\ket{\tilde{\psi}(t)}$ embedding the probability distribution of the random variable $X(t)$.
We explicitly state for the $L(t)$ operator of the VQS and decompose it by the polynomial number of the sum of easily-implementable unitaries.

\subsection{Embedding the probability distribution into a quantum state}
To simulate the trinomial tree model of the target SDE by VQS, we define an unnormalized quantum state containing the discretized probability distribution of the random variable $X(t_j)$:
\begin{equation} \label{eq: direct embedding}
    \ket{\tilde{\psi}(t)} \equiv \sum_{i=0}^{N_x} \mr{Prob}[X(t)=x_i] \ket{i},
\end{equation}
where $\{ \ket{i} \}_{i=0}^{N_x}$ is the computational basis. We call this state a directly embedded state. For simplicity, we assume that $N_x=2^n-1$, where $n$ is the number of qubits.

Note that this embedding of the probability distribution into the quantum state differs from most of the literature, in which (aiming for a quantum advantage) the expectation values of a probability distribution are calculated using QAE~\cite{Brassard2002}.
In the literature, the probability distribution is expressed as a normalized quantum state
\begin{equation} \label{eq: square-root embedding}
    \ket{\psi_\mr{sqrt}} \equiv \sum_i \sqrt{\mr{Prob}[X(t_j)=x_i]} \ket{i}.
\end{equation}
The expectation value of the distribution, $E[f(X(t_j))] \equiv \sum_i f(x_i) \mr{Prob}[X(t_j)=x_i]$ for some function $f$, is computed by the QAE.
In this embedding method, VQS cannot be used because the differential equation describing the time evolution of the probability distribution is nonlinear.
There are ways to solve the nonlinear differential equation with a quantum algorithm~\cite{Lubasch2020, Kyriienko2020, Liu2020, Lloyd2020}, but they require more complicated quantum circuits.

Because our embedding~\eqref{eq: direct embedding} differs from this embedding scheme, we also developed a method for evaluating its expectation values (see Sec.~\ref{sec: expectation}).
Note that the normalization constant $\alpha$ in Eq.~\eqref{eq: unnoraml state def} may be exponentially small. In fact, for a uniform distribution $\mathrm{Prob}[X(t_j)=x_i]=1/2^n$, the normalization constant is $2^{-n/2}$.

\subsection{Reformulating the trinomial tree model and applying the variational quantum simulation}
In the trinomial tree model, the probability $\mr{Prob}[X(t_{j+1}) = x_i]$ is calculated as
\begin{eqnarray}
    \mr{Prob}[X(t_{j+1}) = x_i]
    &=& p_{u}(x_{i-1},t_j) \mr{Prob}[X(t_j) = x_{i-1}] \nonumber \\
    &+& p_{d}(x_{i+1},t_j) \mr{Prob}[X(t_j) = x_{i+1}] \nonumber \\
    &+& p_{m}(x_i,t_j) \mr{Prob}[X(t_j) = x_i].
\end{eqnarray}
Substituting the transition probabilities~\eqref{eq: p_u},~\eqref{eq: p_d} and \eqref{eq: p_m} into this expression and denoting $P(x, t) \equiv \mr{Prob}[X(t) = x]$, we get
\begin{eqnarray}
    &&\frac{P(x_i, t_{j+1}) - P(x_i,t_j)}{\Delta t} \nonumber\\
    &=&  \frac{1}{2}\left(\frac{\sigma^2(x_{i-1},t_j)}{\Delta x^2}+\frac{\mu(x_{i-1},t_j)}{\Delta x}\right) P(x_{i-1},t_j) \nonumber\\
    &+& \frac{1}{2}\left(\frac{\sigma^2(x_{i+1},t_j)}{\Delta x^2}-\frac{\mu(x_{i+1} ,t_j)}{\Delta x}\right) P(x_{i+1},t_j) \nonumber\\
    &-& \frac{\sigma^2(x_i,t_j)}{\Delta x^2} P(x_i,t_j).
\end{eqnarray}
In the limit $\Delta t \rightarrow 0$, one obtains
\begin{eqnarray}
    &&\frac{d \vec{P}(t)}{d t} = L(t)\vec{P}(t), \label{eq: master} \\
    &&(L(t))_{i,k} = \left\{
    \begin{array}{ll}
        \frac{1}{2}\left(\frac{\sigma^2(x_k,t)}{\Delta x^2}+\frac{\mu(x_k,t)}{\Delta x}\right) & (i = k + 1)    \\
        \frac{1}{2}\left(\frac{\sigma^2(x_k,t)}{\Delta x^2}-\frac{\mu(x_k,t)}{\Delta x}\right) & (i = k - 1)    \\
        - \frac{\sigma^2(x_k,t)}{\Delta x^2}                                                   & (i = k)        \\
        0                                                                                      & \mr{otherwise}
    \end{array}
    \right., \nonumber \\
\end{eqnarray}
where $\vec{P}(t) \equiv (P(x_0,t),P(x_1, t), \dots, P(x_{2^n-1}, t))^T$.

As shown in Eq.~\eqref{eq: master}, the time evolution of the state $\ket{\tilde{\psi}(t)}$, or
\begin{eqnarray}
    \frac{d}{dt} \ket{\tilde{\psi}(t)} = \hat{L}(t) \ket{\tilde{\psi}(t)},  \label{eq: master in qs}
\end{eqnarray}
where
\begin{eqnarray}
    \hat{L}(t) \equiv \sum_{i,k=0}^{2^n-1} (L(t))_{i,k} \ket{i}\!\bra{k},  \label{eq: def of L(t)}
\end{eqnarray}
corresponds to the time evolution of the probability distribution $\{ \mr{Prob}[X(t)=x_i] \}_{i=0}^{2^n-1}$.
Equation~\eqref{eq: master in qs} is the essence of our proposal to simulate VQS-based SDE simulation: specifically, the VQS algorithm applied to Eq.~\eqref{eq: master in qs} obtains the time-evolved probability distribution as the quantum state $\ket{\tilde{\psi}(t)}$.
Hereafter, when the distinction is clear in context, we denote the operator $\hat{L}(t)$ by $L(t)$ as in Eq.~\eqref{eq: master}.

\subsection{Construction of $L(t)$}
As explained in the previous section, in the VQS, we evaluate Eqs.~\eqref{eq: M matrix} and \eqref{eq: V vector}, and decomposes $L(t)$ into a sum of easily-implementable unitaries (composed of single-qubit, two-qubit, and few-qubit gates). These evaluations are important for a feasible VQS.
This subsection discusses the explicit decomposition of $L(t)$ given by Eq.~\eqref{eq: def of L(t)}.

To express the operator $L(t)$ in Eq.~\eqref{eq: def of L(t)}, we define operators
\begin{equation}
    V_+(n) \equiv \sum_{i=0}^{2^n-2} \ket{i+1}\!\bra{i},
    V_-(n) \equiv \sum_{i=1}^{2^n-1} \ket{i-1}\!\bra{i}.
\end{equation}
These operators can be constructed from the $n$-qubit cyclic increment/decrement operator
\begin{equation}
    \mr{CycInc}(n) \equiv \sum_{i=0}^{2^n-1} \ket{i+1}\!\bra{i}, \:
    \mr{CycDec}(n) \equiv \sum_{i=0}^{2^n-1} \ket{i-1}\!\bra{i},
\end{equation}
where $\ket{-1}, \ket{2^n}$ are identified with $\ket{2^n-1}, \ket{0}$, respectively.
These gates are implemented as a product of $O(n)$ Toffoli, CNOT, and X gates with $O(n)$ ancilla qubits~\cite{Li2014}.
$V_+(n)$($V_-(n)$) is constructed from $\mr{CycInc}(n)$($\mr{CycDec}(n)$) and an $n$-qubit-control $Z$ gate $C^nZ \equiv \sum_{i=0}^{2^n-2} \ket{i}\!\bra{i} - \ket{2^n-1}\!\bra{2^n-1}$,
which can be implemented~\cite{Nielsen2010} as a product of $O(n^2)$ Toffoli, CNOT, and single qubit gates.
Using $\frac{1}{2}\left(C^nZ+I^{\otimes n}\right) = \sum_{i=0}^{2^n-2} \ket{i}\!\bra{i}$, we can show that
\begin{eqnarray}
    V_+(n) &=&\mathrm{CycInc}(n) \cdot \frac{1}{2}\left(C^nZ+I^{\otimes n}\right), \\
    V_-(n) &=& \frac{1}{2} \left(C^nZ + I^{\otimes n}\right) \cdot \mathrm{CycDec}(n),
\end{eqnarray}
meaning that $V_{\pm}(n)$ can be decomposed into a sum of two unitaries composed of $O(n^2)$ few-qubit gates.
Finally, we define the operator $D(n)$ by
\begin{eqnarray}\label{eq: def of D}
    D(n)
    &=& \sum_{i=0}^{2^n-1} i \ket{i}\!\bra{i} \nonumber \\
    &=& \frac{2^n-1}{2}I^{\otimes n} - \sum_{i=1}^n 2^{n-i-1} Z_i, \label{eq: diag}
\end{eqnarray}
where $Z_i$ is a $Z$ gate acting on the $i$th qubit.
Therefore, $D(n)$ is a sum of $O(n)$ unitaries composed of a single-qubit gate.
It follows that
\begin{eqnarray}
    V_+(n) (D(n))^m &=&
    \sum_{i=0}^{2^n-2} i^m \ket{i+1}\!\bra{i}, \\
    V_-(n)(D(n))^m &=&
    \sum_{i=1}^{2^n-1} i^m  \ket{i-1}\!\bra{i}.
\end{eqnarray}

Let us recall
\begin{eqnarray}
    L(t) &=& \sum_{i=0}^{2^n-2} \frac{1}{2}\left(\frac{\sigma^2(x_i,t)}{\Delta x^2} + \frac{\mu(x_i,t)}{\Delta x}\right) \ket{i+1}\!\bra{i} \nonumber \\
    &+&\sum_{i=1}^{2^n-1} \frac{1}{2}\left(\frac{\sigma^2(x_i,t)}{\Delta x^2} - \frac{\mu(x_i,t)}{\Delta x}\right) \ket{i-1}\!\bra{i} \nonumber \\
    &-& \sum_{i=0}^{2^n-1} \frac{\sigma^2(x_i,t)}{\Delta x^2} \ket{i}\!\bra{i}. \nonumber
\end{eqnarray}
Expanding $\sigma^2(x_i, t)$ and $\mu(x_i, t)$ as
\begin{equation}\label{eq: extraction of mu,sigma}
    \sigma^2(x_i, t) = \sum_{m=0}^{m_\sigma} a_{\sigma,m}(t) x_i^m,
    \mu(x_i, t) = \sum_{m=0}^{m_\mu} a_{\mu,m}(t) x_i^m,
\end{equation}
we can decompose $L(t)$ as follows:
\begin{eqnarray}
    && L(t)  \nonumber\\
    &=& \sum_{m=0}^{m_\sigma} a_{\sigma,m}(t) (\Delta x)^{m-2} \left(\frac{V_+(n)+V_-(n)}{2} - I \right)(D(n))^m \nonumber\\
    &+& \sum_{m=0}^{m_\mu} a_{\mu, m}(t)(\Delta x)^{m-1} \left( \frac{V_+(n)-V_-(n)}{2} - I \right)(D(n))^m. \nonumber
\end{eqnarray}
$V_+(n)(D(n))^m$, $V_-(n)(D(n))^m$ and $(D(n))^m$ are composed of the sum of $O(n^m)$ unitaries, each composed of $O(n^2)$ few-qubit gates.
In typical SDEs, the orders $m_\sigma, m_\mu$ can be set to small values. For example, geometric Brownian motion case, $m=1$ (see Sec.~\ref{sec: numerics}).
Therefore, the $L(t)$ decomposition realizes a feasible VQE (Eq.~\eqref{eq: master in qs}).

\section{Calculation of Expectation Values \label{sec: expectation}}
In the previous section, we propose a method to simulate the SDE by calculating the dynamics of the probability distribution of a random variable $X(t)$ using VQS.
However, in many cases, the goal of the SDE simulation is not the probability distribution of $X(t)$, but the expectation value $E[f(X(t))]$ of $X(t)$ for some function $f$.
In this section, we introduce a means of calculating this expectation value.

\subsection{Problem Setting}
Given a function $f(x): \mathbb{R} \to \mathbb{R}$, we try to calculate the expectation value $E[f(X(T))]$ of the SDE~\eqref{eq: general SDE def} at time $t=T$.
The expectation value can be explicitly written as
\begin{equation} \label{eq: expectation}
    E[f(X(T))] \equiv \sum_{i=0}^{2^n-1} f(x_i) \mr{Prob}[X(T)=x_i].
\end{equation}
Here, we assume that $f(x)$ in the interval $[a_k, a_{k+1}] \in \{ [0, a_1], [a_1, a_2], \ldots, [a_{d-1}, x_\mr{max}] \} \: (k=0,\ldots,d-1)$ is well approximated by $L$th order polynomials $f_k(x) = \sum_{m=0}^{L} a_m^{(k)} x^m$.
The additional error from this piecewise polynomial approximation is evaluated in Appendix~\ref{app: error from piecewise polynomial approximation}.
As $x$ is finite, the range of $f$ is also finite.
Thus, by shifting the function $f$ by a constant, we can ensure that the range of $f$ is positive and that the expectation value is also positive, i.e. $E[f(X(T))] \geq 0$.
In most situations (such as pricing of European call options as we see in Sec.~\ref{subsection: pricing european call option}) the number of intervals $d$ does not scale with the number of qubits $n$.

\subsection{General formula for calculating expectation values\label{subsection: general formula}}
We now on compute the expectation value~\eqref{eq: expectation} using the quantum state $\ket{\tilde{\psi}(t)}$ (Eq.~\eqref{eq: direct embedding}).
First, we consider a non-unitary operator satisfying
\begin{equation} \label{eq: def of Sf}
    S_f\ket{0} = \sum_{i=0}^{2^n-1} f(x_i) \ket{i}
\end{equation}
and decompose $S_f$ into a sum of easily-implementable unitaries as $S_f = \sum_i \xi_i Q_i$ with complex coefficients $\xi_i$.
It follows that
\begin{equation} \label{eq: expectation_formula}
    \bra{\tilde{\psi}(t)} \left( S_f\ket{0}\!\bra{0} S_f^\dag \right) \ket{\tilde{\psi}(t)} =  \left( E[f(X(T))] \right)^2.
\end{equation}
As $\ket{0}\!\bra{0} = I - C^nZ \cdot X^{\otimes n}$ is also a sum of easily-implementable unitaries as explained in the previous subsection, the Hermitian observable $S_f\ket{0}\!\bra{0} S_f^\dag$ is decomposed as
\begin{equation} \label{eq: decomposition of observable}
    S_f\ket{0}\!\bra{0} S_f^\dag = \sum_{i,i'} \xi_i \xi_{i'}^* \left( Q_i Q_{i'}^\dag - Q_i (C^nZ \cdot X^{\otimes n}) Q_{i'}^\dag \right),
\end{equation}
which is again a sum of unitaries.
With this decomposition, the left-hand side of Eq.~\eqref{eq: expectation_formula} is computed by evaluating $\braket{\tilde{\psi}(t)|Q_i Q_{i'}^\dag|\tilde{\psi}(t)}, \braket{\tilde{\psi}(t)|Q_i (C^nZ \cdot X^{\otimes n}) Q_{i'}^\dag|\tilde{\psi}(t)}$.
Because we set $E[f(X(T))] \geq 0$, the left hand side of Eq.~\eqref{eq: expectation_formula} will determine the expectation value.

\begin{figure}
    \centering
    \mbox{
    \Qcircuit @C=2.0em @R=1.4em {
    \lstick{\ket{0}} &  \gate{H} & \ctrl{1}  & \gate{H}     &  \meter\\ \lstick{\ket{\tilde{\psi}(t)}} &  \qw {/} & \gate{U} &   \qw & \qw \\
    }
    }
    \caption{Quantum circuit for evaluating the real part of an expectation value $\Re\bra{\tilde{\psi}(t)}U\ket{\tilde{\psi}(t)}$ of a unitary operator $U=Q_{i}Q_{i'}^\dagger,Q_{i}C^nZ\cdot X^{\otimes n}Q_{i'}^\dagger$. The imaginary part of the expectation value $\Im\bra{\tilde{\psi}(t)}U\ket{\tilde{\psi}(t)}$ is evaluated by the circuit with an $S^\dag$ gate inserted to the left of the second $H$ gate.}
    \label{fig: hadamard_test}
\end{figure}
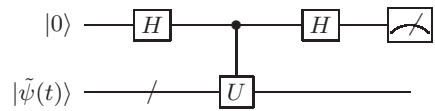

There are two options to evaluate the quantities $\braket{\tilde{\psi}(t)|Q_i Q_{i'}^\dag|\tilde{\psi}(t)}$ and $\braket{\tilde{\psi}(t)|Q_i (C^nZ \cdot X^{\otimes n}) Q_{i'}^\dag|\tilde{\psi}(t)}$.
The first one is to use the Hadamard test depicted in Fig.~\ref{fig: hadamard_test}. The second one is to use quantum phase estimation~\cite{Knill2007, Wang2019}. The former one requires shallower quantum circuits but is inefficient in terms of the number of measurements to determine the quantities with fixed precision.
The detailed computational complexity of these methods is given in Sec.~\ref{sec: discussion} and Appendix~\ref{app: complexity of expectation}.

Next, we explain the construction of the operator $S_f$ in Eq.~\eqref{eq: def of Sf} and its decomposition.
We first define an operator
\begin{equation}
    S_{\chi_{[0, a]}} \ket{0} = \sum_{i=0}^{2^n-1}\chi_{[0,a]}(x_i)\ket{i} = \sum_{x_i\in[0,a]} \ket{i},
    \label{eq: def of indicator}
\end{equation}
where $\chi_{[0,a]}(x)$ is the indicator function valued as $1$ for $x \in [0, a]$ and $0$ for otherwise.
Using the binary expansion of $a/\Delta x$, we can obtain the decomposition of $S_{\chi[0,a]}$ hence the decomposition of $S_f$.
As $a\in[0,x_{\mathrm{max}}]$, there exists $k_a\in\mathbb{N}$ such that $\Delta x 2^{k_a-1} \leq a < \Delta x 2^{k_a}, 0 < k_a \leq n$.
The binary expansion of $a/\Delta x$ is given by $a/\Delta x=\sum_{j=0}^{k_a-1}s_j2^j,s_j\in\{0,1\}$. We define the list of $l$ as $l_1,l_2,\dots,l_B(=k_a-1)$ satisfying $s_l=1$ in ascending order, and also define an interval
\begin{eqnarray}
    \chi^a_l &=& \left[2^{l_B}+\sum_{j=0}^{l-1}s_{j}2^{j}+1, 2^{l_B}+\sum_{j=0}^{l}s_{j}2^{j}\right]
\end{eqnarray}
for $l\in\{l_1,l_2,\dots,l_B\}$.
Using $\chi_l^a$, we devide $[0,a/\Delta x]$ into disjoint intervals as follows:
\begin{equation}
    [0,a/\Delta x] = [0,2^{l_B}]\cup\chi_{l_1}^a\cup\dots\cup\chi_{l_B}^a.
\end{equation}
The indicator operator $S_{\chi_{[0,a]}}$ is obtained by summing the indicator operators on each interval.
In binary expansion, the $k_a$th and the $l$th bit of $i\in\chi_l^a$ are $1$, and the bit below $l$ is either $0$ or $1$. Accordingly, $X$ should act on the bit taking $1$, and $H$ should act on the bit taking either of $\{0,1\}$. The indicator operator $S_{\chi_l^a}$ on $\chi_l^a$ is defined as follows:
\begin{eqnarray}
    &&S_{\chi^a_l}\ket{0}=\ket{0}^{\otimes k_a-1}\otimes\ket{1}\bigotimes_{j=0}^{n-k_a-l-1}\ket{s_{n-k_a-j}}\otimes\left(\sum_{j=0}^{l}\left| j \right>\right)\nonumber\\
    &&=2^{l/2}I^{\otimes k_a-1}\otimes X\bigotimes_{j=0}^{n-k_a-l-1}\mathbb{X}_{s_{n-k_a-j}}\otimes H^{\otimes l}\ket{0},
    \label{eq: indicator on interval chi}
\end{eqnarray}
where
\begin{eqnarray}
    \mathbb{X}_{s} = \left\{
    \begin{array}{cc}
        X & (s=1) \\
        I & (s=0)
    \end{array}.
    \right.
\end{eqnarray}
In addition, we define
\begin{equation}
    S_{\chi_{[0,2^{k_a-1}]}}\ket{0}=2^{(k_a-1)/2}I^{\otimes n-k_a+1}H^{\otimes k_a-1}\ket{0}.
    \label{eq: indicator on first interval}
\end{equation}
We can construct $S_{\chi_{[0,a]}}$ by summing Eqs.~\eqref{eq: indicator on interval chi} and \eqref{eq: indicator on first interval} on each interval. $S_{\chi_{\alpha_k}}$ on interval $\alpha_k \equiv [a_k,a_{k+1}]$ is
\begin{eqnarray}
    S_{\chi_{\alpha_k}}=S_{\chi_{[0,a_{k+1}]}}-S_{\chi_{[0,a_{k}]}},
\end{eqnarray}
which is a sum of at most $O(n)$ unitaries composed of $O(n)$ gates.
Using $S_{\chi_{\alpha_k}}$, we obtain
\begin{eqnarray}\label{eq: S_f}
    S_f\ket{0}=\sum_{k=0}^{d-1}\sum_{x_i\in{{\alpha_k}}}f(x_i)S_{\chi_{\alpha_k}}\ket{0},
\end{eqnarray}
and $S_f$ is constructed as
\begin{eqnarray}\label{eq: decomposition of Sf}
    S_f=\sum_{k=0}^{d-1}\sum_{m=0}^{L}a_m^{(k)}(D(n))^mS_{\chi_{\alpha_k}}.
\end{eqnarray}

In summary, evaluation of the expectation value is calculated by the following steps.
\begin{enumerate}
    \item Divide the domain of the target function $[0,x_{\mathrm{max}}]$ into intervals $[a_k, a_{k+1}] \in \{ [0, a_1], [a_1, a_2], \ldots, [a_{d-1}, x_\mr{max}] \}$.
    \item Approximate the function in each interval $[a_k,a_{k+1}]$ by Eq.~\eqref{eq: decomposition of Sf}.
    \item Decompose $S_f\ket{0}\bra{0}S_f^\dagger$ into a sum of unitary terms and calculate each term using the circuits in Fig.~\ref{fig: hadamard_test}.
\end{enumerate}
As $S_{\chi_{[a_k,a_{k+1}]}}, (D(n))^m, \ket{0}\bra{0}$ is the sum of $O(n)$, $O(n^m)$ and $O(1)$ unitaries composed of $O(n)$, $O(1)$ and $O(n^2)$ gates, respectively, $S_f\ket{0}\bra{0}S_f^\dagger$ is the sum of $O(d^2n^{2L+2})$ unitaries and each $Q_i$ is composed of at most $O(n^{4})$ gates. 

When the target function $f$ on each interval is written by a low-degree polynomial (i.e., $L$ is small), especially by a linear function (as in the pricing of European call options shown below), our algorithm can efficiently calculate the expectation value because the number of unitaries $O(d^2n^{2L+2})$ gets not so large.
When the function $f$ is approximated by the polynomial, we can estimate the error of the expectation value stemming from that approximation.
If we want to suppress the error below $\epsilon$, the number of unitaries becomes $O(x_{\max}^2\epsilon^{-\frac{2}{L+1}}n^{2L+2})$ (the derivation is presented in Appendix~\ref{app: error from piecewise polynomial approximation}).
Note that as $L$ is increased, $\epsilon^{-\frac{2}{L+1}}$ becomes smaller while $n^{2L+2}$ becomes larger. 
The number of unitaries, therefore, is not monotonic with respect to $L$, and there may be an optimal $L$ for the desired accuracy.
We note that evaluation of expectation values of those unitaries can be performed completely in parallel by independent quantum devices.

\subsection{Pricing of The European Call Option\label{subsection: pricing european call option}}
As a concrete example, we present the pricing of a European call option with the Black-Scholes (BS) model, which is one of the simplest financial derivatives.
The holder of a European call option is entitled to buy the asset at a predetermined strike price at maturity.
The price of a European call option with strike price $K\geq0$, interest rate $r\geq0$, and maturity $T\geq0$ is defined by the conditional probability
\begin{eqnarray}
    e^{-rT}E_Q\left[ \max(X_T-K, 0)\middle| X_0 =x_0 \right].
    \label{eq: european option}
\end{eqnarray}
Here, $E_Q$ denotes the expectation value under the risk-neutral probability measure.
Stochastic processes are assumed to follow geometric Brownian motion in the BS model, but are described by more complex mechanisms in other models.
Even in these models, the expression Eq.~\eqref{eq: european option} of the price of the European call option is the same with the present case.

Setting the probability distribution of $X_T$ conditioned by $X_0=x_0$ as $\left\{\mathrm{Prob}\left[X_T=x_i\middle|X_0=x_0\right]\right\}_{i=0}^{2^n-1}$, the expectation value is
\begin{eqnarray}
    &&e^{-rT}E_Q\left[ \max(X_T-K, 0)\middle| X_0=x_0 \right] \nonumber\\
    &=& e^{-rT}\sum_{i=0}^{2^n-1}\mathrm{Prob}\left[X_T=x_i\middle| X_0=x_0\right]\max(x_i-K, 0). \nonumber\\
\end{eqnarray}
For simplicity, we assume $\Delta x=1$ and $K=2^k<2^n-1, k\in\mathbb{N}$. We thus obtain
\begin{equation}\label{eq: S_f for european call option}
    S_{\max(i-K,0)}=(D(n)-KI)S_{\chi_{[K,2^{n-1}]}}.
\end{equation}
In this case, there are only two intervals $[0,K-1]$ and $[K,2^{n-1}]$, and the polynomial in each interval is of first-order degree at most.
Therefore, we can calculate the price of the European call option by Eq.~\eqref{eq: expectation_formula}.
\section{Possible Advantages of Our Method \label{sec: discussion}}
In this section, we discuss the advantages of our method compared to previous studies, as well as the possible quantum advantages.

In general, the SDEs addressed in this paper can be transformed into a partial differential equation (PDE) of the function $e_f(x, t)$, where $e_f(x,t)$ gives the expectation value $E[f(X(T-t))|X(0)=x]$, by Feynman-Kac formula~\cite{Shreve2004}.
In fact, the authors of~\cite{Fontanela2019} performed a variational quantum computation of a PDE of this function.
We point out two advantages of our method compared with this strategy using Feynman-Kac formula.
First, the resulting PDE by Feynman-Kac formula must be solved backwardly in time from $t=T$ to $t=0$, with the initial condition at $t=T$ being related to the functional form of $f(X)$.
It is not trivial to prepare the initial state $\ket{\psi(T)}$ corresponding to the initial condition; the authors of \cite{Fontanela2019} executed an additional VQE to prepare the initial state.
Second, when using the Feynman-Kac formula, the initial condition of the PDE is different for each function $f$ for which we want to calculate the expectation value $E[f(X(T))]$. 
If we want to calculate a different expectation value $E[f'(X(T))]$, we need to run the whole algorithm simulating the PDE with the different initial state corresponding to $f'$.
On the other hand, in our method, once we perform VQS, we obtain the probability distribution of $X(T)$ as a quantum state and the corresponding variational parameters to reproduce it. 
We only need to redo the part of the expectation value calculation (Sec.~\ref{sec: expectation}) for different $f'$.

The authors of~\cite{Rebentrost2018} embedded the probability distribution by quantum arithmetic.
Their embedding, proposed in~\cite{Grover2002}, requires $O(2^n)$ gates to embed the probability distribution into an $n$-qubit quantum state.
To moderate the gate complexity, the authors of~\cite{Zoufal2019} embedded the probability distribution using a quantum generative adversarial network, which requires only $O(\mathrm{Poly}(n))$ gates. The probability distribution function can also be approximated by a $l$th-order piecewise polynomial, which can be embedded with $O(ln^2)$ gates even in quantum arithmetic \cite{Haner2018}.
However, both methods require prior knowledge of the probability distribution to be embedded.
In contrast, our method does not require prior knowledge of the embedding probability distribution since our method simulates the time evolution of a given SDE.

We now compare the computational cost to calculate expectation values with previous studies.
In \cite{Rebentrost2018} and \cite{Zoufal2019}, by employing QAE, the expectation value (Eq.~\eqref{eq: expectation}) was calculated by using an oracle that is complex quantum gate reflecting the functional form of $f$ for $O(1/\epsilon)$ times, where $\epsilon$ is the precision for the expectation values.
The classical Monte Carlo method requires $O(1/\epsilon^2)$ sampling for precision $\epsilon$, so their methods provide a second-order acceleration.
On the other hand, our method measures the expectation value of each term of Eq.~\eqref{eq: decomposition of observable} using the Hadamard test (Fig.~\ref{fig: hadamard_test}) or the quantum phase estimation (QPE)~\cite{Knill2007, Wang2019}.
As shown in Appendix~\ref{app: complexity of expectation},
the total number of measurements to obtain the expectation value with precision $\epsilon$ is $O(1/\gamma\epsilon^2)$ for the Hadamard test and $O(\log(1/\gamma\epsilon))$ for QPE, where $\gamma$ is some factor.
We note that the depth of the circuit is $O(1/\gamma\epsilon)$ in QPE, which is in the same order as the QAE whereas our method requires not an complicated oracle but a relatively-small unitary.
Hence, when the factor $\gamma$ is not too small, our method combined with QPE can also exhibit quantum advantage for the evaluation of the expectation values.
The factor $\gamma$ depends on the parameters of the polynomial approximation $(a_k^{(m)}, d, L)$, the domain of the approximated function $x_{\mr{max}}$, and the probability distribution $\{\mr{Prob}[X=x_i]\}_{i=0}^{2^n-1}$. 
The detailed evaluation of $\gamma$ is described in Appendix~\ref{app: complexity of expectation}.%
\section{Numerical Results\label{sec: numerics}}
In this section, our algorithm is applied to two stochastic processes, namely, geometric Brownian motion and an Ornstein-Uhlenbeck process, which are commonly assumed in financial engineering problems.
Geometric Brownian motion simply models the fluctuations of asset prices, and the Ornstein-Uhlenbeck process is a popular model of interest rates.

\subsection{Models}
Geometric Brownian motion is equivalent to setting $\mu(X(t), t)=rX(t), \sigma(X(t),t)=\sigma X(t)$ in Eq.~\eqref{eq: general SDE def}, where $r$ and $ \sigma$ are positive constants.

The Ornstein–Uhlenbeck process is equivalent to setting $\mu(X(t),t)=-\eta(X(t)-r), \sigma(X(t),t)=\sigma$ in Eq.~\eqref{eq: general SDE def}, where $\eta$, $r$ and $ \sigma>0$ are constants.

\begin{figure}
    \begin{minipage}{\columnwidth}
        \mbox{
        \Qcircuit @C=1em @R=.7em {
        \lstick{\ket{0}} & \gate{RY(\theta_1)} & \ctrl{1} & \qw        & \qw       & \targ   & \gate{RY(\theta_{k,1})} & \qw   \\
        \lstick{\ket{0}} & \gate{RY(\theta_2)} &   \targ    & \ctrl{1} & \qw       & \qw    & \gate{RY(\theta_{k,2})}  & \qw  \\
        \lstick{\ket{0}} & \gate{RY(\theta_3)} &   \qw      & \targ    & \ctrl{1}  & \qw   & \gate{RY(\theta_{k,3})}  & \qw     \\
        \lstick{\ket{0}} & \gate{RY(\theta_4)} &   \qw      & \qw    & \targ     & \ctrl{-3} & \gate{RY(\theta_{k,4})} & \qw \gategroup{1}{3}{4}{7}{.7em}{--}      \\
        \\
        &&&& \rstick{ \mathrm{repeat} \ k\  \mathrm{times}}&&&
        \\
        }
        }
    \end{minipage}\\
    \caption[ansatz circuit]{In a depth-$k$ circuit, CNOT and RY gates (enclosed by dashed lines) are repeated $k$-times. The circuit has $4(k+1)$ parameters.
    }\label{fig: ansatz}
\end{figure}
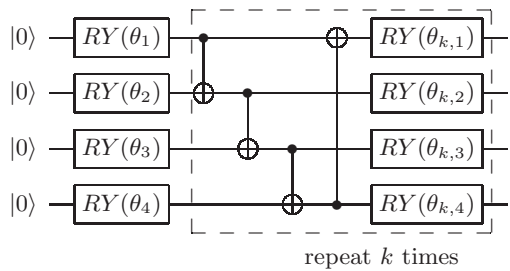

The ansatz circuit is identical for both models and shown in Fig.~\ref{fig: ansatz}. As the amplitudes of the quantum state must be real, the ansatz contains only CNOT and RY gates. This depth-$k$ circuit repeats the entangle blocks composed of CNOTs and RY gates $k$ times.
The parameters of geometric Brownian motion were $r=0.1, \sigma=0.2, \Delta x = 1$, and $t\in[0,4]$ and those of the Ornstein–Uhlenbeck process were $r=7, \sigma=0.5, \eta=0.01, \Delta x = 1$, and $t\in[0,4]$.
We simulate the quantum circuits without noise using numpy~\cite{harris2020} and jax~\cite{jax2018}. We set the number of qubits $n=4$ and the number of repetitions of entangle blocks $k=2,3$.

\subsection{Results}
\begin{figure*}
    \centering
    \subfloat[Dynamics of geometric Brownian process]{
        \centering
        \includegraphics[width=\columnwidth]{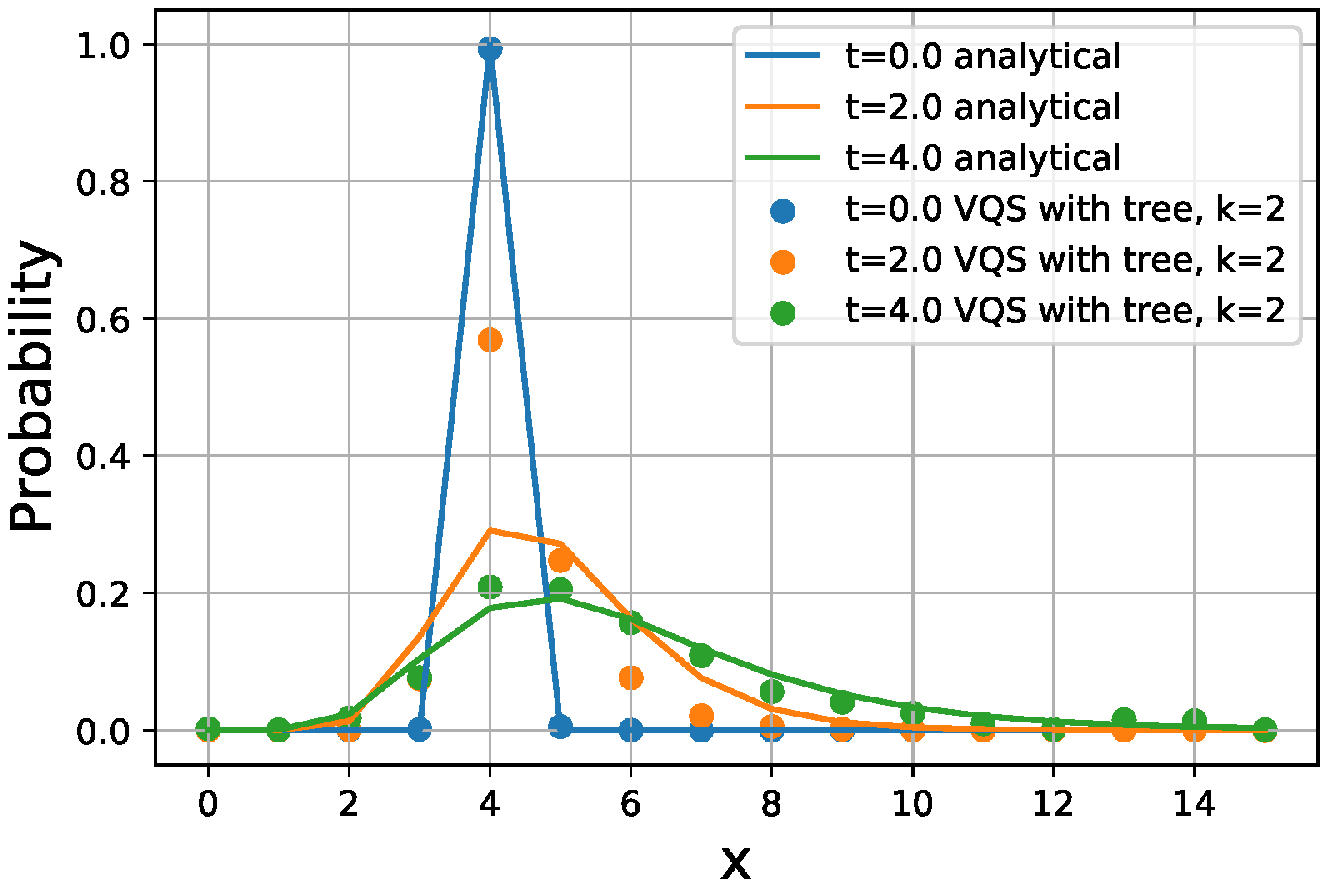}
        \label{fig:dynamics_gbm}
    }
    \subfloat[Dynamics of Ornstein-Uhlenbeck process]{
        \centering
        \includegraphics[width=\columnwidth]{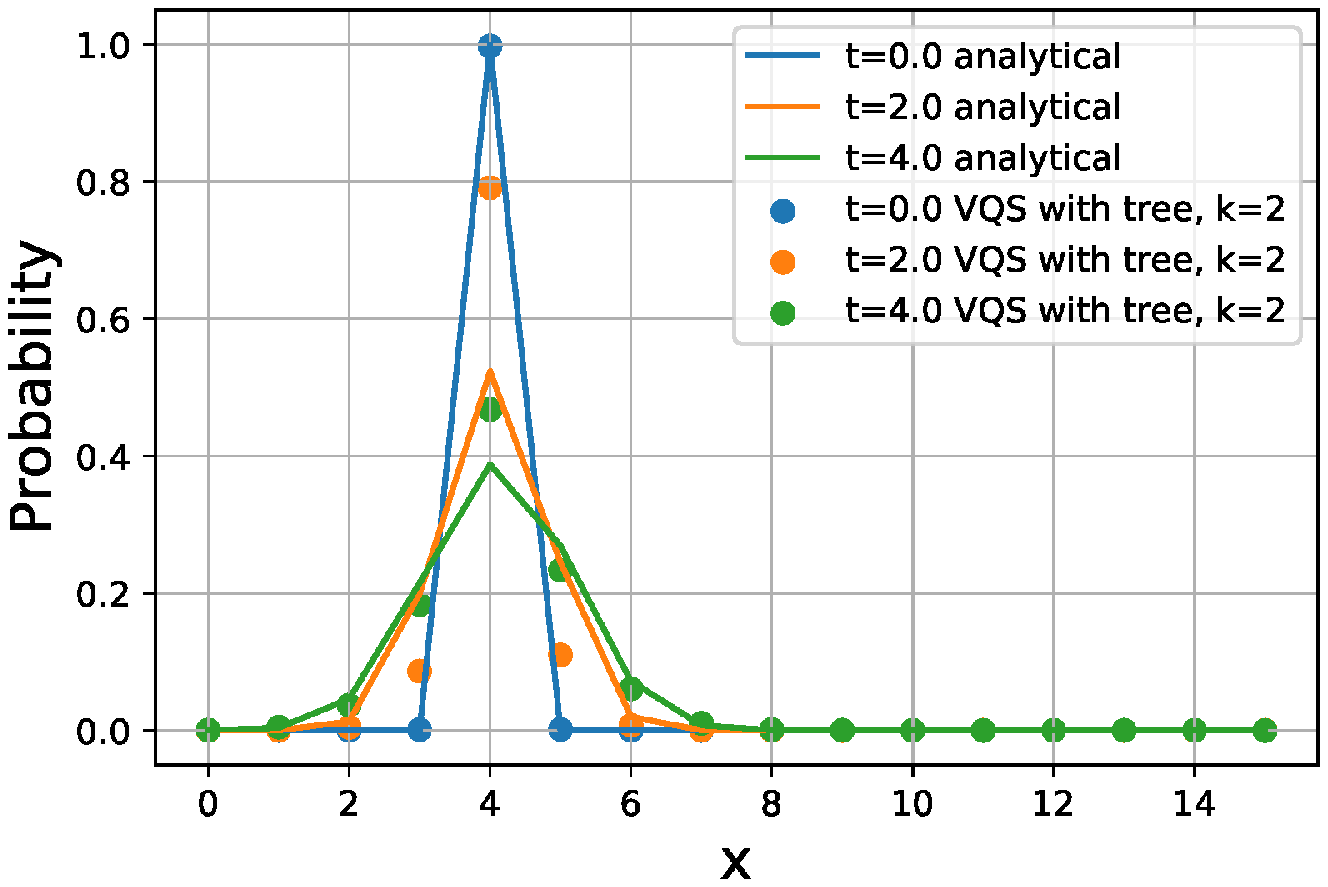}
        \label{fig:dynamics_ou}
    }\\
    \subfloat[Time dependence of mean in geometric Brownian motion]{
        \centering
        \includegraphics[width=\columnwidth]{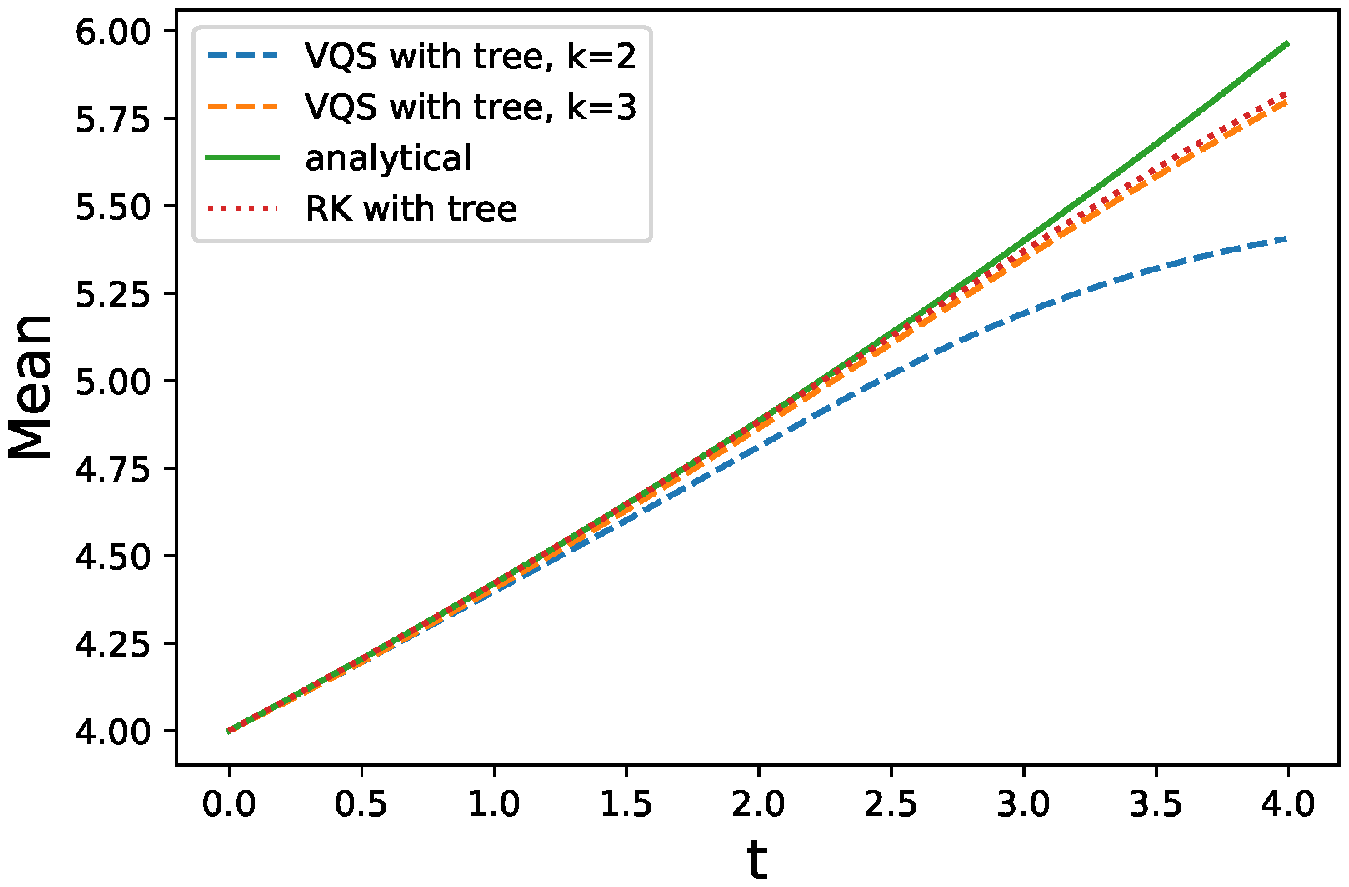}
        \label{fig:mean_gbm}
    }
    \subfloat[Time dependence of mean in Ornstein-Uhlenbeck process]{
        \centering
        \includegraphics[width=\columnwidth]{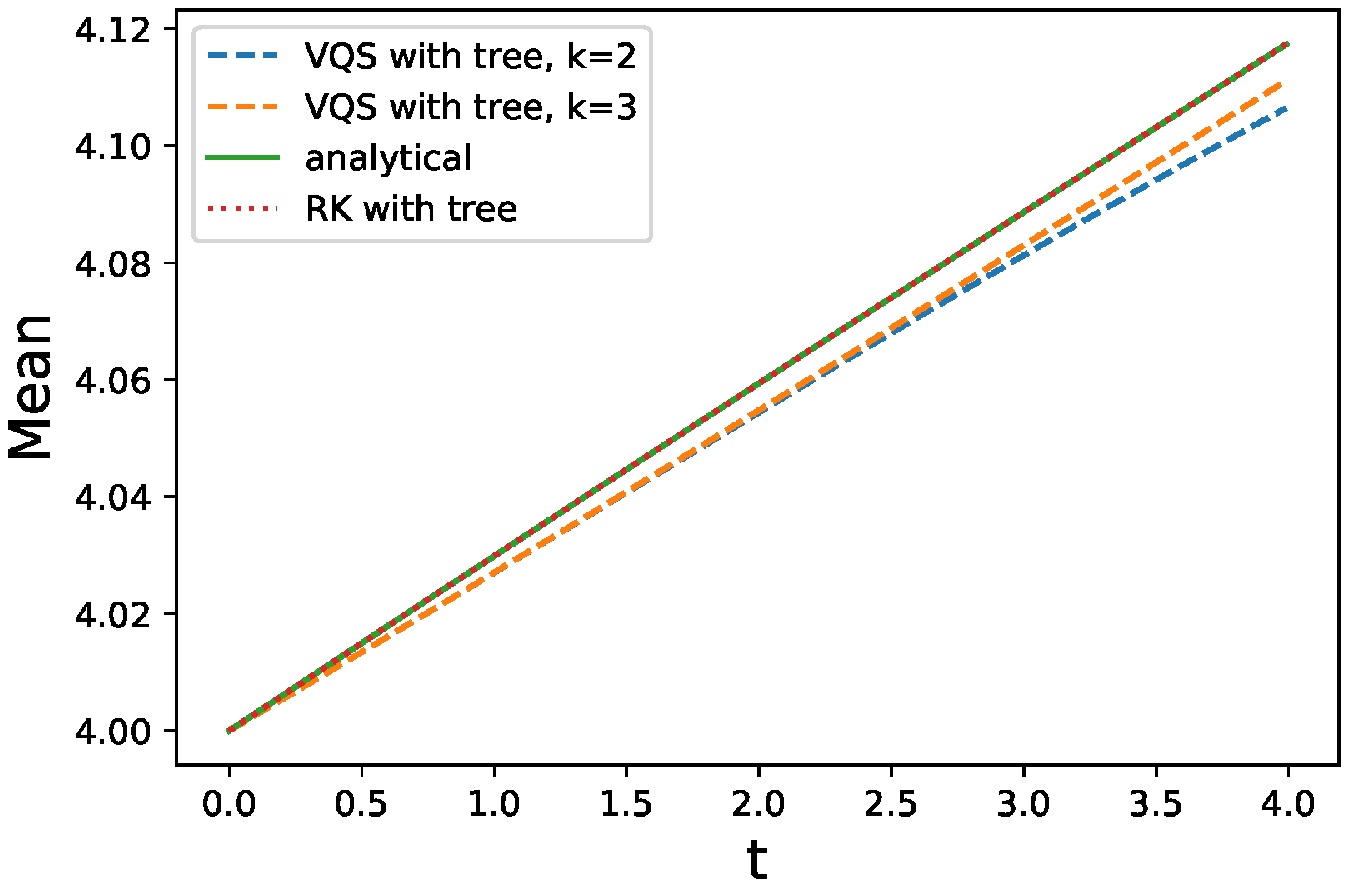}
        \label{fig:mean_ou}
    }\\
    \subfloat[Time dependence of variance in geometric Brownian motion]{
        \centering
        \includegraphics[width=\columnwidth]{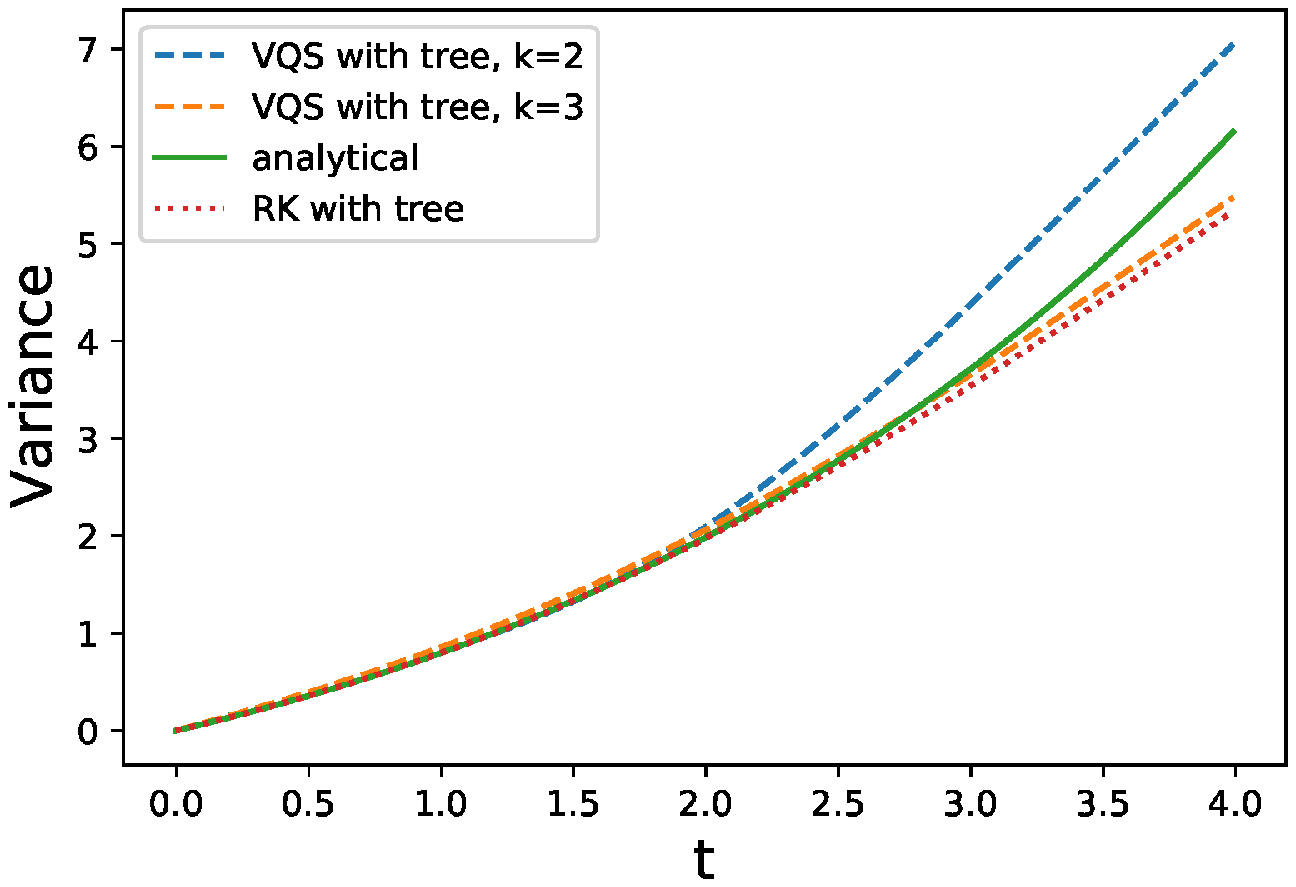}
        \label{fig:var_gbm}
    }
    \subfloat[Time dependence of variance in Ornstein-Uhlenbeck process]{
        \centering
        \includegraphics[width=\columnwidth]{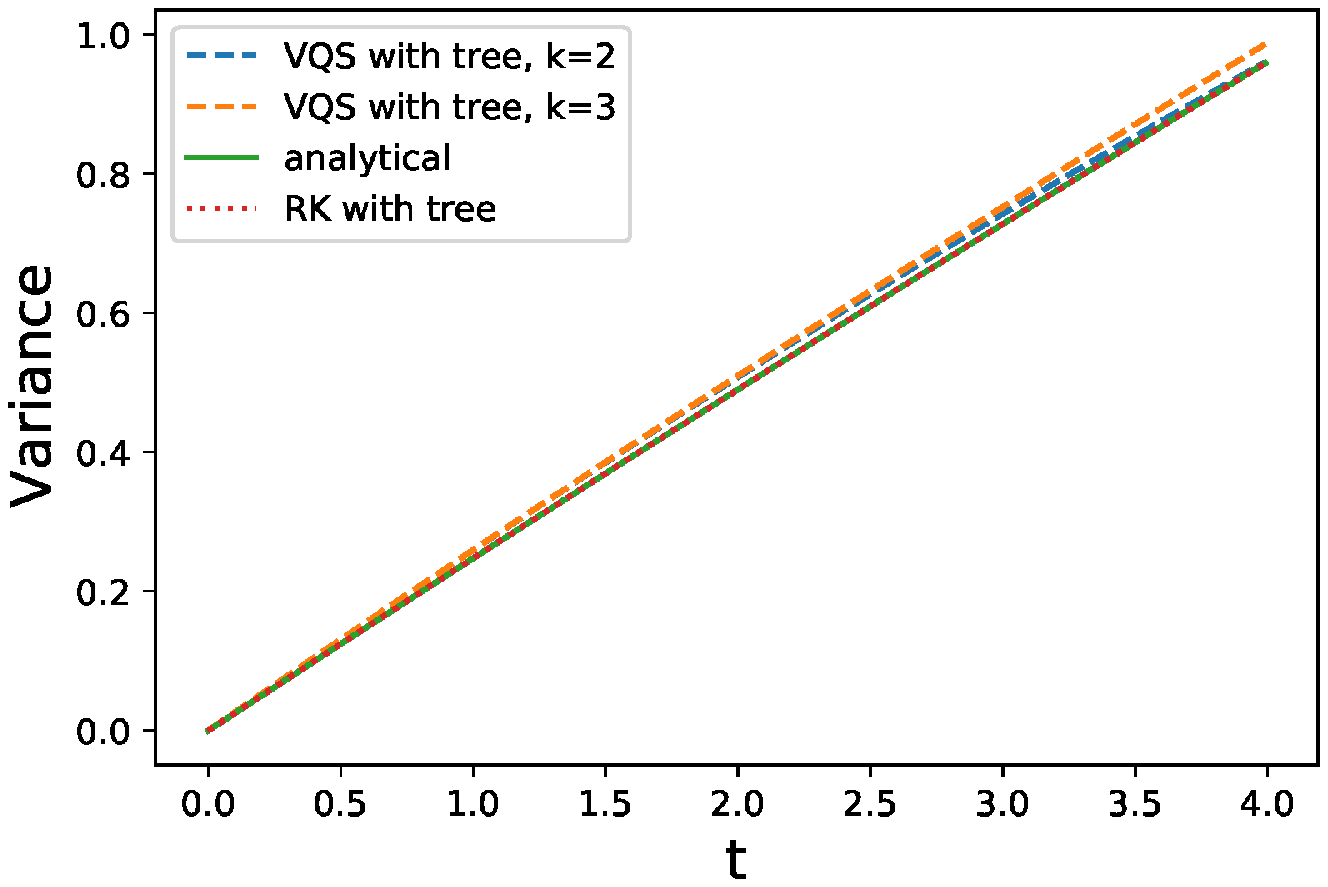}
        \label{fig:var_ou}
    }
    \\
    \caption{
        \protect\subref{fig:dynamics_gbm}, \protect\subref{fig:dynamics_ou}: Exact solutions of the SDE (solid lines) and numerical simulation of our algorithm (circles);
        \protect\subref{fig:mean_gbm}, \protect\subref{fig:mean_ou}: time dependence of the means and \protect\subref{fig:var_gbm}, \protect\subref{fig:var_ou} variances of the encoded probability distributions. Dashed lines show the numerical solutions of VQS with a $k=2,3$ depth ansatz. Dotted lines show the numerical Runge-Kutta solutions of the linear differential equation (Eq.~\eqref{eq: master}). Solid lines show the exact solutions of the SDE.
        \protect\subref{fig:dynamics_gbm}, \protect\subref{fig:mean_gbm}, \protect\subref{fig:var_gbm}: Geometric Brownian motion with parameters $r=0.1, \sigma=0.2$.
        \protect\subref{fig:dynamics_ou}, \protect\subref{fig:mean_ou}, \protect\subref{fig:var_ou}: Ornstein-Uhlenbeck Process with parameters $r=7, \sigma=0.5, \eta=0.01$.
    }
    \label{fig: results}
\end{figure*}

Panels \subref{fig:dynamics_gbm} and \subref{fig:dynamics_ou} of Fig.~\ref{fig: results} present the numerical simulations of geometric Brownian motion and the Ornstein-Uhlenbeck process, respectively.
In comparison, we also provide a probability density function (PDF) for the solution of the SDE equation obtained by solving the Fokker-Planck equation~\cite{Risken1996} analytically.
We can see that our method describes the time evolution of the probability distribution well.

We calculated the means (Fig.~\ref{fig: results}\subref{fig:mean_gbm},\subref{fig:mean_ou}) and variances (Fig.~\ref{fig: results}\subref{fig:var_gbm},\subref{fig:var_ou}) of the resulting distributions.
We also present the mean and variance obtained from the analytical solution and the solution of \eqref{eq: master} using the Runge-Kutta method.
Because of the approximation with the tree model, even the results of the Runge-Kutta method slightly differ from the analytical solution.
In the case of VQS with $k=2$, we see that the error from the analytical solution is larger than that of the $k=3$ case.
This is because the number of VQS parameters is less than the number of lattice points in the event space when $k=2$, i.e., the degrees of freedom of ansatz are less than the degrees of freedom of the system, and thus the errors due to the ansatz appear.
In the case of $k=3$, the number of parameters in ansatz is sufficient, and thus the results are closer to the results of the Runge-Kutta method.

\section{Conclusion \label{sec: conclusion}}
This paper proposed a quantum-classical hybrid algorithm that simulates SDEs based on VQS.
A continuous stochastic process was discretized in a trinomial tree model and was reformulated as a linear differential equation. The obtained differential equation was solved with VQS, obtaining quantum states representing the probability distribution of the stochastic processes. As our method can embed the probability distribution of the solution of a given SDE into the quantum state, it is applicable to general SDEs.
We note that our methods can apply to the Fokker-Plank equation, which also gives the time-evolution of the probability distributions of SDE solutions.

Because the embedding methods of the probability distribution differ in the proposed method and the conventional quantum algorithm, we proposed another method for computing the expectation value.
We approximated the functions to calculate expectation values by piecewise polynomials and constructed operators corresponding to the polynomial in each interval.
The operators were constructed as sums of unitary operators, which are composed of easily-implementable gates. The expectation value was then computed using the sum of unitary operators.
Our algorithm was validated in classical simulations of geometric Brownian motion and the Ornstein-Uhlenbeck process. Both processes were well simulated by the algorithm. Our algorithm is expected to efficiently simulate other stochastic processes, provided that $L(t)$ can be written as a polynomial linear combination of unitary matrices.

Let us summarize the computational cost of our method presented in this work.
Our method consists of two parts; one is to perform VQS to simulate the SDE, and the other is to calculate the expectation value of the SDE solution.
In the part of running VQS, we decompose matrix $L(t)$ in Eq.~\eqref{eq: master} into a sum of $O(n^{m_{\max}})$ different unitaries composed of $O(n^2)$ few-qubit gates, where $m_{\max}$ is the largest order of the polynomial expansion of $\mu,\sigma$ in Eq.~\eqref{eq: extraction of mu,sigma}.
At each time step of VQS, the vector $V_k$ in Eq.~\eqref{eq: V vector} is evaluated as a sum of $O(n^{m_{\max}})$ measurement results of the circuits depicted in Fig.~\ref{fig:circuitMV}.
As $m_{\max}$ is typically finite and small ($\sim 1,2$) in most practical applications, the computational cost (i.e., the number of gates in quantum circuits, the number of different circuits to run) of the simulation of SDE is $O(\mathrm{Poly}(n))$.
In contrast, QLSA~\cite{Harrow09,Berry2014,Berry2017} requires much deeper and more complex quantum circuits and a large number of ancilla qubits because it uses the Hamiltonian simulation and the quantum Fourier transform.
This is an advantage of our method leveraging the variational quantum algorithm. 

In the part of the expectation value evaluation of the SDE solution, we evaluate it by running different $O(d^2n^{2L+2})$ quantum circuits, where $d$ and $L$ are the number of intervals and the order of the piecewise polynomial approximation of the function $f$ in Eq.~\eqref{eq: expectation}, respectively.
Each circuit is constructed to compute an expectation value $\braket{\psi|U|\psi}$ of a unitary $U$ that contains $O(n^4)$ quantum gates.
When we adopt the Hadamard test (Fig.~\ref{fig: hadamard_test}) as such a quantum circuit, the number of measurements to suppress statistical error of the expectation value below $\epsilon$ is $O(1/\gamma \epsilon^2)$, where $\gamma$ is a factor defined in Appendix~\ref{app: complexity of expectation}.
This $O(1/\epsilon^2)$ scaling is the same as the classical Monte Carlo method to compute the expectation values from the probability distribution of the SDE solution.
When we choose the QPE-type circuit to evaluate $\braket{\psi|U|\psi}$, the number of measurements becomes $O(\log(1/\gamma\epsilon))$ while the depth of the circuit in terms of $U$ is $O(1/\gamma\epsilon)$.
This situation can provide a quantum advantage for computing the expectation value of the SDE solution.
The error from the piecewise polynomial approximation of $f$ can be made small by increasing $d$ or $L$, which is detailed in Appendix~\ref{app: error from piecewise polynomial approximation}.

This study focused on computational finance because financial engineering is among the most popular applications of stochastic processes. Pricing of derivatives, and many other problems in financial engineering, satisfy the conditions of the proposed method.
However, as the stochastic processes themselves are quite general, the proposed method is expected to contribute to solving problems in various fields.

\begin{acknowledgments}
    The authors thank Keisuke Fujii for valuable discussions on variational quantum algorithms.
\end{acknowledgments}

\bibliographystyle{apsrev4-1}
\bibliography{bibliography}

\appendix
\section{Complexity of calculating expectation value\label{app: complexity of expectation}}

This Appendix derives the computational complexity of calculating the expectation by Eq.~\eqref{eq: expectation_formula}.
To limit the error $\epsilon$ in expectation value $E=\sqrt{\bra{\tilde{\psi}}S_f\ket{0}\bra{0}S_f^\dag\ket{\tilde{\psi}}}$, we show the upper limit of error $\epsilon'$ of the expectation value for each term in Eq.~\eqref{eq: decomposition of observable}, and find the number of measurements and gate complexity required to achieve this error.
We assume that $S_f \left| 0\right> \left<0\right| S_f^\dag$ can be written as a linear combination of $N_U$ unitary operators as follows
\begin{eqnarray}\label{eq: decomposition of Sf by unitary}
    S_f \left| 0\right> \left<0\right| S_f^\dag = \sum_{i=1}^{N_U} \beta_i U_i,
\end{eqnarray}
where $\{U_i\}$ are unitary operators.
We denote the error of expectation values $\bra{\psi}U_i\ket{\psi}$, where $\ket{\psi}$ is the normalized state,
\begin{align}
\ket{\psi}=\frac{1}{\sqrt{\sum_{j=0}^{2^n-1}p_j^2}}\ket{\tilde{\psi}},
\end{align}
where $p_j=\mr{Prob}[X(t)=x_i]$.
We define the error as $\epsilon'$ of the expectation value of each term in a state $\ket{\psi}$.
That is, the estimated expectation value of each term $\tilde{u}_i$ satisfies
\begin{eqnarray}
    \left|\tilde{u}_{i}-\left<\psi\right|U_i\left|\psi\right>\right|\leq\epsilon'.
\end{eqnarray}
The error in the linear combination of expectation values is determined as
\begin{eqnarray}
    \left|\sum_{i=1}^{N_U} \beta_i \tilde{u}_i -\bra{\psi} \sum_{i=1}^{N_U} \beta_iU_i \ket{\psi}\right| &=&\left|\sum_{i=1}^{N_U} \beta_i \left(\tilde{u}_i -\bra{\psi} U_i\ket{\psi}\right)\right| \nonumber\\
    &\leq&\sum_{i=1}^{N_U}\left|\beta_i\right|\left| (\tilde{u}_{i} -\bra{\psi} U_i \ket{\psi})\right| \nonumber\\
    &\leq&\epsilon'\sum_{i=1}^{N_U} \left|\beta_i\right|.
\end{eqnarray}
Denoting the estimation of $E$ as $\tilde{E}$, we have
\begin{align}
    \left|\tilde{E}-E\right|&\leq\sum_{j=0}^{2^n-1}p_j^2\frac{\epsilon'\sum_{i=1}^{N_U}\left|\beta_i\right|}{\tilde{E}+E}
    \nonumber\\&\sim \sum_{j=0}^{2^n-1}p_j^2\frac{\epsilon'\sum_{i=1}^{N_U}\left|\beta_i\right|}{2E}.
\end{align}
To upper limit the error $\epsilon$ in $E$, the error $\epsilon'$ must satisfy following condition:
\begin{equation}
    \epsilon'\lesssim\frac{2\epsilon E}{\sum_{j=0}^{2^n-1}p_j^2\sum_{i=1}^{N_U}\left|\beta_i\right|}=\gamma\epsilon,
\end{equation}
where $\gamma\equiv\left( 2E/\sum_{j=0}^{2^n-1}p_j^2\cdot\sum_{i=1}^{N_U}\left|\beta_i\right|\right)$.

The Hadamard test (Fig.~\ref{fig: hadamard_test}) requires $O(1/\epsilon'^2)$ measurements to limit the error in the expectation value to $\epsilon'$, and the depth of the quantum circuit is $O(1)$ in terms of the unitary $U (=Q_{i}Q_{i'}^\dagger,Q_{i}C^nZ\cdot X^{\otimes n}Q_{i'}^\dagger)$ except for the circuit to prepare the quantum state.
On the other hand, the number of measurements to achieve the same accuracy with QPE is $O(\log(1/\epsilon'))$, but the depth of the quantum circuit in terms of the unitary $U$ is $O(1/\epsilon')$~\cite{Knill2007,Wang2019}.
The total number of measurements is equal to the number of measurements for each term multiplied by $N_U$.
Note that $N_U=O(d^2n^{2L+2})$ with $L$th-order piecewise polynomial approximation of the function $f$ with $d$ intervals (see Sec.~\ref{subsection: general formula}).
Therefore, the total number of measurements required to calculate the expectation value is $O(d^2n^{2L+2}/\gamma\epsilon^2)$ by Hadamard test and $O(d^2n^{2L+2}\log(1/\gamma\epsilon))$ by QPE.
QPE requires extra $U$ gates, the number of which is $O(1/\gamma\epsilon)$.

Finally, we provide the estimation of the value of $\gamma$ as follows.
The factor $\sum_{j=0}^{2^n-1}p_j^2$ satisfies $\sum_{j=0}^{2^n-1}p_j^2\leq1$ and then $\gamma\geq2E/\sum_{i=1}^{N_U}\left|\beta_i\right|$.
To evaluate $\sum_{i=1}^{N_U}\left|\beta_i\right|$, we use Eq.~\eqref{eq: decomposition of observable} and obtain 
\begin{align}
    \sum_{i=1}^{N_U}|\beta_i|=2\sum_{l, l'}|\xi_l\xi_{l'}^*|= 2\sum_{l,l'}|\xi_l||\xi_{l'}|.
\end{align}
We estimate the upper limit of sum of absolute values of coefficients in Eq.~\eqref{eq: decomposition of Sf} to evaluate $\sum_l|\xi_l|$.
From Eq.~\eqref{eq: def of D}, the absolute values of coefficients of $(D(n))^m$ is at most $O(2^{nm})=O(x_{\max}^m)$.
As $S_{\chi_{\alpha_{k}}}$ is a linear combination of $S_{\chi_{l}^\alpha}$, the absolute values of coefficients of $S_{\chi_{\alpha_{k}}}$ is at most $O(2^{(k_a-1)/2})$ from Eq.~\eqref{eq: indicator on first interval}.
Since $k_a$ satisfies $0<k_a\leq n$ by definition, $O(2^{(k_a-1)/2}) = O(\sqrt{x_{\max}})$.
The largest $|\xi_l|$ is as large as $\left(\max_k\max_m\left(|a_m^{(k)}|x_{\max}^m\right)\right)O\left(\sqrt{x_{\max}}\right)$ from Eq.~\eqref{eq: decomposition of Sf}.
Thus, we obtain
\begin{align}\label{eq: evaluation of beta}
    \sum_{i=1}^{N_U}|\beta_i| \lesssim O\left(\left[\max_k\max_m|a_m^{(k)}|x_{\max}^m\right]^2 d^2 n^{2L+2} x_{\max}\right).
\end{align}
Therefore, $\gamma$ is larger than the ratio of $E$ and the right-hand side of Eq.~\eqref{eq: evaluation of beta}.
\section{Multivariate stochastic differential equation\label{app: multi-variables}}
In this Appendix, we apply our proposed method to an SDE with multiple variables.

\subsection{Definition and construction of the tree-model approximation}
Let us consider a SDE with $D$ variables,
\begin{equation} \label{eq: multi SDE}
    dX_d(t) = \mu_d(X_d, t) dt + \sigma_d(X_d, t)dW_d,
\end{equation}
for $X_1(t), \dots, X_D(t)$, where $\{ W_d \}_{d=1}^D$ describes the Brownian motion with correlation $\mr{Corr}[W_k, W_l] = \rho_{kl}$.
For simplicity, we assume an event space of each variable $X_d(t)$ as $[0, x_\mr{max}^{(d)}]$ , and divide it into $N_x + 1$ points; that is, $x^{(d)}_i \equiv i\Delta x^{(d)}, \Delta x^{(d)} \equiv x_\mr{max}^{(d)}/N_x$ ($d=1,\ldots,D$).
The time period of the simulation, $t \in [0, T]$, is divided into $N_t + 1$ points, $t_j \equiv j\Delta t$; that is $\Delta t \equiv T/N_t$.

We define a lattice of the tree model for Eq.~\eqref{eq: multi SDE} with nodes $(i_1, \ldots, i_D; j)$ representing the random variables $(X_1(t_j), \ldots, X_D(t_j)) = (x_{i_1}^{(1)}, \ldots, x_{i_D}^{(D)})$, where $i_d = 0,\ldots,N_x, \: j=0,\ldots,N_t, \: d=1,\ldots,D$.
The node transitions during time $t_j \to t_{j+1}$ are of three types:
\begin{eqnarray*}
    (1) \: (i_1, \ldots, i_D; j) &\to& (i_1, \ldots, i_D; j+1), \\
    (2) \: (i_1, \ldots, i_D; j) &\to& (i_1, \ldots, i_k \pm 1, \ldots, i_D; j+1), \\
    (3) \: (i_1, \ldots, i_D; j) &\to&  (i_1, \ldots, i_k \pm 1, \ldots, i_l \pm 1, \ldots, i_D; j+1),
\end{eqnarray*}
where $1 \leq k < l \leq D$.
Type (1), (2) and (3) transitions occur to nodes with identical variable values, to nodes where one-variable $X_k$ hops to its adjacent values, and to nodes where two variables ($X_k$ and $X_l$) hop to their adjacent values, respectively.
The transition probabilities associated with type (1), (2) and (3) transitions are respectively given by
\begin{eqnarray*}
    p_m(x_{i_1}^{(1)}, \ldots, x_{i_D}^{(D)}, t), \\
    p_{u,d}^{(k)}(x_{i_1}^{(1)}, \ldots, x_{i_D}^{(D)}, t), \\
    p_{uu,ud,du,dd}^{(k,l)}(x_{i_1}^{(1)}, \ldots, x_{i_D}^{(D)}, t),
\end{eqnarray*}
where the subscript $u(d)$ corresponds to the sign $+(-)$.

The transition probabilities can be determined identically to those of the one-variable SDE.
The SDE~\eqref{eq: multi SDE} at at $(X_1(t_j),\ldots,X_D(t_j)) = (x_{i_1}^{(1)},\ldots,x_{i_D}^{(D)})$ is discretized as
\begin{equation}
    X_d(t_{j+1}) - X_d(t_j) = \mu_d(X_d(t_j), t) \Delta t + \sigma_d(X_d(t_j), t) \sqrt{\Delta t} z_d,
\end{equation}
where $\{z_d\}_{d=1}^D$ is sampled from the multi-variable Gaussian distribution, $\mr{E}[z_d]=0, \mr{Var}[z_d]=1, \mr{Corr}[z_k, z_k] = \rho_{kl}$.
The first and second first and second conditional moments satisfy
\begin{eqnarray}\label{eq: E of multi EM}
    E[X_d(t_{j+1}) - X_d(t_j)|X_d(t_j)=x] &=& \mu_d(x,t_j) \Delta t, \nonumber\\
\end{eqnarray}
\begin{eqnarray}\label{eq: Var of multi EM}
    \mr{Var}[X_d(t_{j+1}) - X_d(t_j)|X_d(t_j)=x] &=& \sigma^2_d(x,t_j) \Delta t  \nonumber\\
\end{eqnarray}
for $d=1,\ldots,D$ and the covariance of the variables satisfies
\begin{equation}\label{eq: Cov of multi EM}
    \begin{split}
        \mr{Cov} & [X_k(t_{j+1}) - X_k(t_j), X_l(t_{j+1}) - X_l(t_j) \\
                & |X_k(t_j)=x, X_l(t_j)=y] \\
        = & \sigma_k(x,t_j) \sigma_l(y,t_j) \rho_{kl} \Delta t
    \end{split}
\end{equation}
for $1 \leq k < l \leq D$.
The corresponding quantities in the tree model are
\begin{eqnarray}\label{eq: E of multi tree}
    &&E[X_d(t_{j+1}) - X_d(t_j)|X_d(t_j)=x] \nonumber\\
    &=& \left(
    p_u^{(d)} - p_d^{(d)}
    + \sum_{k=1}^{d-1} \left( p_{uu}^{(k,d)} - p_{ud}^{(k,d)} + p_{du}^{(k,d)} - p_{dd}^{(k,d)} \right)
    \right. \nonumber\\
    &&\left.
    + \sum_{l=d+1}^D \left( p_{uu}^{(d,l)} + p_{ud}^{(d,l)} - p_{du}^{(d,l)} - p_{dd}^{(d,l)} \right)
    \right) \Delta x^{(d)}
\end{eqnarray}
\begin{eqnarray}\label{eq: Var of multi tree}
    &&\mr{Var}[X_d(t_{j+1}) - X_d(t_j)|X_d(t_j)=x] \nonumber\\
    &=& \left(
    p_u^{(d)} + p_d^{(d)}
    + \sum_{k=1}^{d-1} \left( p_{uu}^{(k,d)} + p_{ud}^{(k,d)} + p_{du}^{(k,d)} + p_{dd}^{(k,d)} \right)
    \right. \nonumber\\
    &&\left.
    + \sum_{l=d+1}^D \left( p_{uu}^{(d,l)} + p_{ud}^{(d,l)} + p_{du}^{(d,l)} + p_{dd}^{(d,l)} \right)
    \right) \left( \Delta x^{(d)} \right)^2
\end{eqnarray}
for $d=1,\ldots,D$ and
\begin{equation}\label{eq: Cov of multi tree}
    \begin{split}
        \mr{Cov} & [X_k(t_{j+1}) - X_k(t_j), X_l(t_{j+1}) - X_l(t_j) \\
                & |X_k(t_j)=x, X_l(t_j)=y] \\
        = & \left(p_{uu}^{(k,l)} - p_{ud}^{(k,l)} - p_{du}^{(k,l)} + p_{dd}^{(k,l)}\right) \Delta x^{(k)} \Delta x^{(l)}.
    \end{split}
\end{equation}

As is the same for the case of a single variable we set the transition amplitudes by equating Eqs.~\eqref{eq: E of multi EM},\eqref{eq: Var of multi EM},\eqref{eq: Cov of multi EM} with \eqref{eq: E of multi tree},\eqref{eq: Var of multi tree},\eqref{eq: Cov of multi tree}.
If the solutions of $p_{u,d}^{(k)},p_{uu, ud,du,dd}^{(k,l)}$ are proportional to $\Delta t$, the linear differential equation can be derived by taking the limit of $\Delta t\rightarrow0$ (as in the one-dimensional case Eq.~\eqref{eq: master}).

When $D>1$, one should note the numbers of variables and conditional expressions.
As the numbers of $p_m$, $p_{u,d}^{(k)}, p_{uu, ud,du,dd}^{(k,l)}$ are $1, 2D, 2D(D-1)$, respectively, the number of independent variables is $2D^2$ under the normalized probability conditions.
On the other hand, the number of equations of the mean, variance, and covariance are $D,D,D(D-1)/2$, respectively, so the total number of equations is $D(D+3)/2$. When $D>1$, the number of variables exceeds the number of conditions, so an infinite number of transition probabilities satisfy the condition.

Here, we show there is indeed a solution of the transition amplitudes which admit taking limit $\Delta t\rightarrow 0$ and obtain the linear differential equitation of the probability distributions of the SDE. Fixing $p_{dd}^{(k)}=p_{ud}^{(k)}=p_{du}^{(k)}=0$, the number of variables becomes $D(D+3)/2$, which is slightly asymmetric (because we consider only $p_{uu}^{k}$ to be nonzero), but agrees with the number of conditional expressions.
In this case, the transition probabilities are
\begin{eqnarray}
    p_{uu}^{(k,l)} &=& \frac{\sigma_k \sigma_l \rho_{kl}}{\Delta x^{(k)} \Delta x^{(l)}}\Delta t \label{eq: multi p_uu},\\
    p_{u}^{(d)} &=& \frac{1}{2}\left( \frac{\sigma^2_d}{\left(\Delta x^{(d)}\right)^2}+\frac{\mu_d}{\Delta x^{(d)}} \right)-\sum_{k\neq d} \frac{\sigma_k \sigma_d \rho_{kd}}{\Delta x^{(k)} \Delta x^{(d)}},\nonumber\\ \label{eq: multi p_u} \\
    p_{d}^{(d)}&=&\frac{1}{2}\left(\frac{\sigma^2_d}{\left(\Delta x^{(d)}\right)^2}-\frac{\mu_d}{\Delta x^{(d)}} \right),\label{eq: multi p_d}\\
    p_m&=&1-\left[\sum_{d=1}^D \left( \frac{\sigma^2_d}{\left(\Delta x^{(d)}\right)^2} - \sum_{k\neq d} \frac{\sigma_k \sigma_d \rho_{kd}}{\Delta x^{(k)} \Delta x^{(d)}} \right) \right.\nonumber\\
        &&- \left.\sum_{k\neq l}\frac{\sigma_k\sigma_l \rho_{kl}}{\Delta x^{(k)} \Delta x^{(l)}}\right]\Delta t \nonumber\\
    &=&1-\sum_{d=1}^D \frac{\sigma^2_d}{\left(\Delta x^{(d)}\right)^2}\Delta t.
    \label{eq: multi p_m}
\end{eqnarray}
Here, we omit the arguments of $\mu_d$ and $\sigma_d$ to simplify the notation.

\subsection{Mapping to VQS and construction of $L(t)$}
In the multivariate case, we can construct $L(t)$ as described in Sec.~\ref{sec: sde_by_vqs}.
For notational simplicity, we denote
$\ket{i_1,\dots, i_{D} } = \ket{\boldsymbol{i}}$,$\ket{ i_1,\dots,i_{d-1}, i_d\pm1, i_{d+1},\dots, i_{D}}=\ket{ \boldsymbol{i}\pm\boldsymbol{e}_d }$,$\ket{ i_1,\dots, i_k+1 ,\dots,i_l+1,\dots i_{D} }=\ket{ \boldsymbol{i}+\boldsymbol{e}_k+\boldsymbol{e}_l }.$
Using Eqs.~\eqref{eq: multi p_uu}~\eqref{eq: multi p_u}~\eqref{eq: multi p_d} and~\eqref{eq: multi p_m}, we obtain
\begin{eqnarray}
    L(t)&=&\frac{1}{2}\sum_{d=1}^{D}\sum_{i_d=0}^{2^n-2}\sum_{i_{-d}}\left[ \frac{\sigma^2_d}{\left(\Delta x^{(d)}\right)^2}+\frac{\mu_d}{\Delta x^{(d)}} \right. \nonumber\\
        &&\left.-\sum_{k\neq d} \frac{\sigma_k \sigma_d \rho_{kd}}{\Delta x^{(k)} \Delta x^{(d)}}\right]\ket{ \boldsymbol{i}+\boldsymbol{e}_d}\bra{\boldsymbol{i}} \nonumber\\
    &+&\frac{1}{2}\sum_{d=1}^{D}\sum_{i_d=1}^{2^n-1}\sum_{i_{-d}} \left(\frac{\sigma^2_d}{\left(\Delta x^{(d)}\right)^2}-\frac{\mu_d}{\Delta x^{(d)}} \right) \nonumber\\
    &&\times\ket{\boldsymbol{i}-\boldsymbol{e}_d}\bra{\boldsymbol{i}}  \nonumber\\
    &+&\sum_{k\neq l}\sum_{i_{k,l}=0}^{2^n-2}\sum_{i_{-k},i_{-l}}\frac{\sigma_k \sigma_l \rho_{kl}}{\Delta x^{(k)} \Delta x^{(l)}} \nonumber\\
    &&\times\ket{ \boldsymbol{i}+\boldsymbol{e}_k+\boldsymbol{e}_l}\bra{ \boldsymbol{i}} \nonumber\\
    &-&\sum_{k=1}^D\sum_{\boldsymbol{i}}\frac{\sigma^2_d}{\left(\Delta x^{(d)}\right)^2}\ket{\boldsymbol{i} }\bra{\boldsymbol{i}}
\end{eqnarray}
where $\sum_{\boldsymbol{i}}$ denotes the sum of $i_m\in\{0,\dots,2^n-1\}$ for all $ m\in\{1,\dots,D\}$, $\sum_{i_{-d}}$ is the sum of $i_m\in\{0,\dots,2^n-1\}$ for all $ m\neq d$, and $\sum_{i_{-k,-l}}$ is the sum for $i_m\in\{0,\dots,2^n-1\} $ for all $ m\neq k,l$.

Here, we expand $\sigma_k(x^{(k)},t), \mu_k(x^{(k)},t)$ as
\begin{eqnarray}
    \sigma_k(x^{(k)},t)&=&\sum_{m=0}^{m_{\sigma_k}} a^{(k)}_{\sigma,m}(t)(x^{(k)})^{m},
    \\
    \mu_k(x^{(k)},t)&=&\sum_{m=0}^{m_{\mu_k}} a^{(k)}_{\mu, m}(t)(x^{(k)})^{m}.
\end{eqnarray}
We also define the operators
\begin{eqnarray}
    V_+^{(k)}(n)=I^{\otimes k-1}\otimes V_+(n)\otimes I^{\otimes D-k},\\
    V_-^{(k)}(n)=I^{\otimes k-1}\otimes V_-(n)\otimes I^{\otimes D-k},\\
    D^{(k)}(n)=I^{\otimes k-1}\otimes D(n)\otimes I^{\otimes D-k}.
\end{eqnarray}
These operators satisfy the following equations:
\begin{eqnarray}
    &&V_+^{(k)}(n)(D^{(k)}(n))^{m}=\sum_{i_k=0}^{2^n-2}\sum_{i_{-k}}i_k^{m}\ket{\boldsymbol{i}+\boldsymbol{e}_k}\bra{\boldsymbol{i}}\nonumber\\\\
    &&V_-^{(k)}(n)(D^{(k)}(n))^{m}=\sum_{i_k=1}^{2^n-1}\sum_{i_{-k}}i_k^{m}\ket{\boldsymbol{i}-\boldsymbol{e}_k}\bra{\boldsymbol{i}}\nonumber\\\\
    &&V_+^{(k)}(n)(D^{(k)}(n))^{m_k}V_+^{(l)}(n)(D^{(l)}(n))^{m_l}\nonumber\\
    &=&\sum_{i_k=1}^{2^n-2}\sum_{i_l=1}^{2^n-2}\sum_{i_{-k,-l}}i_k^{m_k}i_l^{m_l}\ket{\boldsymbol{i}+\boldsymbol{e}_k+\boldsymbol{e}_l}\bra{\boldsymbol{i}}.
\end{eqnarray}
Using these operators, we can rewrite $L(t)$ as
\begin{eqnarray}
    L(t)&=&\sum_{d=1}^{D}
    \sum_{m_k=0}^{m_{\sigma_d}}\sum_{m_l=0}^{m_{\sigma_d}}a_{\sigma, m_k}^{(d)}a_{\sigma,m_l}^{(d)}\left(\Delta x^{(d)}\right)^{m_k+m_l-2}\nonumber\\
    &\times&\left(\frac{V_+^{(k)}+V_-^{(k)}}{2}-I\right)(D^{(d)}(n))^{m_k+m_l}\nonumber\\
    &+&\sum_{d=1}^{D}
    \sum_{m_k=0}^{m_{\mu_d}}\sum_{m_l=0}^{m_{\mu_d}}a_{\mu, m_k}^{(d)}a_{\mu,m_l}^{(d)}\left(\Delta x^{(d)}\right)^{m_k+m_l-1}\nonumber\\
    &\times&\left(\frac{V_+^{(k)}-V_-^{(k)}}{2}\right)(D^{(d)}(n))^{m_k+m_l}\nonumber\\
    &+&\sum_{k\neq d}\sum_{m_k=0}^{m_{\sigma_k}}\sum_{m_l=0}^{m_{\sigma_l}}a_{\sigma, m_k}^{(k)}a_{\sigma,m_l}^{(l)}\left(\Delta x^{(k)}\Delta x^{(l)}\right)^{-1}\nonumber\\
    &&\times V_+^{(k)}(n)(D^{(k)}(n))^{m_k}V_+^{(l)}(n)(D^{(l)}(n))^{m_l}.\nonumber\\
    \label{eq: multi L}
\end{eqnarray}
As $V^{(k)}_+(n)(D^{(k)}(n))^m, V^{(k)}_-(n)(D^{(k)}(n))^m$ are the sums of $O(n^m)$ unitaries composed of $O(n^2)$ few-qubit gates, Eq.~\eqref{eq: multi L} is feasible decomposition of $L(t)$.

\subsection{Evaluating the expectation value}
To perform computation of the expectation value, we construct a multivariate indicator operator.
In the $D$ dimensional case, the domain of the function is $\prod_{i=1}^{D}[0,x_\mr{max}^{(i)}]$.
In each dimension, we divide $[0,x_\mr{max}^{(i)}]$ into $d$ intervals $\{[a_{0}^{(i)},a_{1}^{(i)}], ,\dots,[a_{d-1}^{(i)},x_{\mr{max}}^{(i)}]\}$ and obtain $d^D$ regions $I(\{k_i\})=\prod_{i=1}^{D}[a_{k_i}^{(i)},a_{k_i+1}^{(i)}]$.
The indicator operator on $I(\{k_i\})$  is represented by the tensor product of the one-dimensional indicator operator Eq.~(\ref{eq: def of Sf}), i.e.,
\begin{eqnarray}
    S_{\chi_{I(\{k_i\})}} = \bigotimes_{i=1}^D S_{\chi_{[a_{k_i}^{(i)},a_{k_i+1}^{(i)}]}}.
\end{eqnarray}
Thus, we can construct
\begin{eqnarray}
    S_f=\sum_{\{k_i\}}\sum_{m=0}^{m_{k_i}}a_m^{(k_i)}(D(n))^mS_{\chi_{I(\{k_i\})}}.
\end{eqnarray}
Note that $S_f\ket{0}\bra{0}S_f^\dagger$ is the sum of $O(n^{2D(m+1)})$ unitaries and each $Q_k$ in Eq.~\eqref{eq: decomposition of observable} is composed of $O(n^{4})$ gates.
In general, the number of sums grows exponentially with the dimensions.
However, even if the correlations between multivariate stochastic processes are important, exponential growth is inconsequential if the function depends on a small number of random variables.
These issues are not unique to our algorithm.
When calculating expectations using QAE, if the arguments of the function are multidimensional, exponentially greater resources are required to build a multidimensional oracle.

Thus, when the number of sums required to construct $S_f$ is independent of the dimension $D$ of the random variable, our algorithm may be particularly effective.

\section{Error from Piecewise Polynomial Approximation\label{app: error from piecewise polynomial approximation}}
In this section, we evaluate the error of the expectation value $E[f(X(T))]$ from the polynomial approximation of the function $f$.

As in the main text, we divide $[0, x_{\max}]$ into $d$ intervals $\{[0, a_1], [a_1, a_2], \ldots, [a_{d-1}, x_\mr{max}] \}$.
For simplicity, we assume the equally-spaced intervals, so the width of the intervals is $h=x_{\mr{max}}/d$,
We ignore the errors in the probability density function $p(x)$ that come from the tree model approximation of the SDE and the incompleteness of the ansatz of VQS because we focus on the error derived from the piecewise polynomial approximation of $f$.

We define the $L$th order residual term of the Taylor expansion of $f$ around $a_k=kh$ as
\begin{align}
    R_k^L(x)=\frac{1}{(L+1)!}f^{(n)}(c)(x-kh)^{L+1},
\end{align}
where $x\in[a_k, a_{k+1}]=[kh, (k+1)h]$ and $c\in[x, (k+1)h]$.
As $x-kh \leq h$, $R_k^L(x)$ is $O(h^{L+1})$.
When we approximate $f$ on $[a_k, a_{k+1}]$ by the $L$th order Taylor expansion $g^L(x)$, the error of expectation value $E_f = \sum_{i=0}^{2^n-1}f(x_i)p(x_i)$ is
\begin{align*}
    &\left|E_f - E_g\right|\nonumber\\
    &=\left|\sum_{k=1}^{d-1}\int_{kh}^{(k+1)h}f(x)p(x)dx-\sum_{k=1}^{d-1}\int_{kh}^{(k+1)h}g^L(x)p(x)dx\right| \nonumber\\
    &=\left|\sum_{k=1}^{d-1}\int_{kh}^{(k+1)h}R_k^L(x)p(x)dx\right| \nonumber\\
    &\leq \max_k\left[\max_{kh\leq x \leq (k+1)h}\left(\left|R_k^L(x)\right|\right)\right] \cdot \sum_{k'=1}^{d-1}\int_{k'h}^{(k'+1)h}p(x)dx \nonumber\\
    &=\max_k\left[\max_{kh\leq x \leq (k+1)h}\left(\left|R_k^L(x)\right|\right)\right] \nonumber\\
    &=O(h^{L+1})
\end{align*}
To suppress the error below $\epsilon$, it is necessary to set $d>x_{\max}\epsilon^{-\frac{1}{L+1}}$.
From the discussion in Sec. IV, $S_f\ket{0}\bra{0}S_f^\dag$ is the sum of $O(d^2n^{2L+2})$ unitaries.
Thus, we can see that $S_f\ket{0}\bra{0}S_f^\dag$ is the sum of $O(x_{\max}^2\epsilon^{-\frac{2}{L+1}}n^{2L+2})$ unitaries.%

\end{document}